\documentclass[12pt,a4paper]{article}
\usepackage[margin=25mm]{geometry}
\usepackage[T1]{fontenc}
\usepackage{lmodern,amsmath,amssymb,amsthm,mathtools,booktabs,array,tabularx, enumitem,relsize}
\usepackage{bbm}
\usepackage{microtype,enumitem,xcolor,fancyhdr,placeins,needspace}
\usepackage[colorlinks=true,linkcolor=blue!45!black,citecolor=blue!45!black,urlcolor=blue!45!black]{hyperref}
\hypersetup{pdftitle={Cross-Correlation Distribution of a Binary Niho Decimation: Finite-Field Proof},pdfauthor={},pdfsubject={Unit-circle root counts and cross-correlation distributions}}
\numberwithin{equation}{section}
\newtheorem{theorem}{Theorem}[section]
\newtheorem{lemma}[theorem]{Lemma}
\newtheorem{proposition}[theorem]{Proposition}
\newtheorem{corollary}[theorem]{Corollary}
\theoremstyle{definition}\newtheorem{definition}[theorem]{Definition}
\theoremstyle{remark}
\newtheorem{remark}{Remark}

\newtheorem{example}{Example}
\newcommand{\F}{\mathbb F}

\newcommand{\Tr}{\operatorname{Tr}}

\newcommand{\eps}{\varepsilon}

\newcommand{\calD}{\mathcal D}

\setlist{itemsep=3pt,topsep=5pt}
\title{
Correlation distribution for a class of Niho exponents via  self-conjugate-reciprocal polynomials and Kloosterman sums
 }
\author{
Yongbo Xia\thanks{Y. Xia is with  the Department of Mathematics and Statistics, South-Central Minzu University,
Wuhan 430074, China. Email: xia@mail.scuec.edu.cn},
Chunlei Li\thanks{C. Li and T. Helleseth are with the  Department of Informatics, University of
		Bergen, N-5020 Bergen, Norway.  Emails: chunlei.li@uib.no, tor.helleseth@uib.no}, Xiangyong Zeng\thanks{ X. Zeng is with Hubei Key Laboratory of Applied Mathematics, Faculty of Mathematics and Statistics, Hubei
		University, Wuhan, 430062, China. Email: xiangyongzeng@aliyun.com},
and Tor Helleseth$^{\dagger}$}

\date{}
\begin{document}
\maketitle
\begin{abstract}
Let $q=2^m$ for an even $m\geq 2$ and let $d=s(q-1)+1$ with
$4s\equiv5\pmod{q+1}$. In this paper, the correlation
distribution between a binary m-sequence of period
$q^2-1$ and its $d$-decimation is considered. The cross-correlation function takes
seven values $q(i-1)-1$ for $0\leq i\leq 6$, 
and we determine their frequencies 
based on the study of low-degree self-conjugate-reciprocal polynomials and three mixed Kloosterman sums. The correlation distribution is finally expressed in
terms of $K_m(1)$ and $K_m(\omega)$, where $K_m(\cdot)$
is the Kloosterman sum over the finite field $\F_q$
 and $\omega$ is a primitive element of $\F_4$.
\end{abstract}

\section{Introduction}\label{sec:intro}
Let $q$ be a power of two and
$\F_{q^2}$ be the finite field with $q^2$ elements. 
Niho in his thesis~\cite{Niho} intensively studied 
the cross-correlation between binary m-sequences of length $q^2-1$ and their $d$-decimated m-sequences when $d$ is of the form $d=s(q-1)+1$ for certain positive integer $s$. 
It turned out this type of integers involved interesting algebraic structures over finite fields
and they were later referred to as \textit{normalized Niho exponents} in the literature. 
In the last decades researchers have explored their applications in different areas, including low-correlation sequences \cite{Niho,Helleseth1976,HLR2007,Dobbertin,Xia2016,Xia2017}, bent functions \cite{Dobbertin2006Bent,LeanderKholosha2006}, highly nonlinear functions \cite{LiHKT2013Walsh}, permutation polynomials \cite{LiHelleseth2017,BaiXia2017}, cyclic codes~\cite{Charpin2004,LiFengGe2014}, Weil sums \cite{Katz2012,HKL2021}, Kloosterman sums~\cite{Dobbertin}, etc. Interested readers may refer to the survey on Niho exponents and their applications in~\cite{li_zeng_survey_2019}.

When a Niho exponent $d$ is coprime to $q^2-1$, 
the following three problems are equivalent~\cite{Dobbertin}: the value distribution of the cross-correlation function between a binary m-sequence of length $q^2-1$
and its $d$-decimated sequence, the value distribution of the Walsh transform of the power function $x^d$ over the finite field $\F_{q^2}$, and the weight distribution of the cyclic code of length $q^2-1$ associated with $\alpha, \alpha^d$, where $\alpha$ is a primitive element of $\F_{q^2}$. More concretely, these three problems reduce to the value distribution of the following Weil sum
\begin{equation}\label{eq:Weilsum}
    W_d(a) = \sum_{x\in \F_{q^2}} (-1)^{\Tr_{2m}(x^d+ax)},\quad a\in \F_{q^2},
\end{equation} where $\Tr_{2m}(\cdot)$ denotes the trace function from $\F_{q^2}=\F_{2^{2m}}$ to $\F_2$.  Actually the
cross-correlation function $C_d(\tau)$ between a binary m-sequence
of length $q^2-1$ and its $d$-decimated sequence is given by
\begin{equation}\label{eq:corfunc}C_d(\tau)=W_d(\alpha^\tau)-1,\quad 0\le \tau < q^2-1.\end{equation}
To determine the value distribution of $W_d(a)$ for a normalized Niho exponent $d=s(q-1)+1$, one typically transforms this problem to counting roots of an associated polynomial $f_{s,a}(x)=x^{2s-1}+ax^s+\bar{a}x^{s-1}+1$ in the unit circle $U=\{u\in \F_{q^2}: u^{q+1} = 1\}$, as the element $a$ runs through $\F_{q^2}$.
It is worth noting that the problem has been investigated more than fifty years. 
Back to 1972 Niho in his PhD thesis~\cite{Niho} tried different choices of $s$, provided a complete $4$-valued distribution of cross-correlation for $s=2$, and left a lot of problems unsolved. Other values of $s$ were considered later~\cite{Hollmann1}. In 2006 Dobbertin, Felke, Helleseth and Rosendahl~\cite{Dobbertin} considered a general form of Niho exponents 
$d=s(q-1)+1$ with $s\equiv 2^r/(2^r\pm 1)\pmod{q+1}$, which includes all known Niho-type decimations for which the cross-correlation function $C_d(\tau)$ takes $4$ possible values. For $s=3$, the polynomial $f_{s,a}(x)$ is quintic and can have at most $5$ roots in $U$, implying that the cross-correlation function may take $6$ possible values;
the cross-correlation distribution for odd $m$ was determined in~\cite{Dobbertin} via Dickson polynomials and the cases of even $m$ were unsettled. Ten years later Xia, Li, Zeng and Helleseth~\cite{Xia2016} resolved the problem for both even and odd $m$
via an interesting connection to the number of weight-five codewords in Zetterberg codes. For $s=4$, while the polynomial $f_{s,a}(x)$ has degree seven,
Niho in his thesis conjectured that it actually had $0,1,2,3,5$ roots in $U$ for even $m$ based on numerical results; this conjecture wasn't resolved until the work of Helleseth, Katz and Li~\cite{HKL2021} in 2021, where  they connected this problem to 
an orbit pattern of roots of self-conjugate-reciprocal polynomials under a conjugate-reciprocal action; and very recently Xiong and Yan~\cite{XiongYan2026} determined the complete distribution of the cross-correlation function for $s=4$ by a detailed study of the case when $f_{s,a}(x)$ has five roots in $U$. The developments of this research line, which traced back to Niho's work in 1972, indicate that the problem tends to
become more difficult as $s$ increases. More importantly, those developments uncovered some intrinsic connections of Niho-exponents and interesting objects in other branches of discrete mathematics and coding theory~\cite{Dobbertin2006Bent,Charpin2004,Xia2016,XiongLZD2016}.

Let $q=2^m$, where $m\ge 2$ is even.  In this paper, we consider the Niho exponent 
\begin{equation}\label{ourniho}
    d=s(q-1)+1,\quad 4s\equiv5\pmod{q+1}.
\end{equation} The exponent $d$ is coprime to $q^2-1$, and this class of
Niho exponents covers some instances listed in~\cite[Table\,3-3]{Niho}.
For the associated polynomial $f_{s,a}(x)$, one can equivalently consider a hexic polynomial $P_a(x) = x^6 + ax^5 +\bar{a}x+1$,
indicating that the cross-correlation function takes at most $7$ values of the form $q(i-1)-1$ for $0\leq i\leq 6$. To determine the cross-correlation distribution for this Niho
exponent, we start by developing general results on
self-conjugate-reciprocal (SCR) polynomials based on
the work in~\cite{HKL2021}. These polynomials are of the form
$f(x)=\sum_{i=0}^d f_i x^i\in\F_{q^2}[x]$
whose coefficients satisfy $f_i=(f_{d-i})^q$
for $0\le i\le d$. This class includes the key polynomials $f_{s,a}(x)$
as special cases.  In this part we present a general method for transforming SCR polynomials over $\F_{q^2}$ to polynomials over $\F_q$ (see Proposition~\ref{prop:trace-transform}). For quartic SCR polynomials over $\F_{q^2}$, we also provide an alternative reduction to cubic resolvent over $\F_q$ (see Proposition~\ref{prop:resolvent}).
This helps us characterize the cases where $P_a(x)$ has exactly $4$, $5$, or $6$ roots in $U$, which are described mainly in terms of trace conditions on involved parameters. Furthermore, we consider the enumeration of those cases by calculating character sums associated with those trace conditions. 
Based on the correspondence between $(x,y)$ and $(x+y, xy)$ in $\F_{q}^2$, 
we transform all involved character sums and express them in terms of Kloosterman sums $K_m(1)$ and $K_m(\omega)$, where $\omega$ is a primitive element of $\F_4$. Finally the complete correlation distribution is determined and is given as follows.
\begin{theorem}\label{thm:distribution}
Let $q=2^m$ with $m\geq 2$ being even and $\omega$ be a primitive element of $\F_4$.
For the Niho exponent $d=s(q-1)+1$ with $s$ satisfying $4s\equiv5\pmod{q+1}$,
the possible values of the cross-correlation function $C_d(\tau)$ in \eqref{eq:corfunc} are
$q(i-1)-1$, $0\le i\le6$ and their respective frequencies $N_i$ are given by
\[ \left\{\begin{aligned}  
N_0&=\frac{53q^2+\left(52(-1)^{m/2}-135\right)q-8K_m(1)^2-9K_m(1)+9K_m(\omega)^2-10}{144},\\
N_1&=\frac{44q^2+\left(131-88(-1)^{m/2}\right)q+20K_m(1)^2+15K_m(1)-24K_m(\omega)^2-101}{120},\\
N_2&=\frac{3q^2+\left(4(-1)^{m/2}-5\right)q-2K_m(1)^2+K_m(1)+3K_m(\omega)^2}{16},\\
N_3&=\frac{2q^2+\left(4(-1)^{m/2}+9\right)q-2K_m(1)^2-9K_m(1)-7}{36},\\
N_4&=\frac{q^2+\left(4(-1)^{m/2}-11\right)q+4K_m(1)^2+3K_m(1)-3K_m(\omega)^2+2}{48},\\
N_5&=\frac{q+1+K_m(1)}{8},\\
N_6&=\frac{q^2+\left(9-52(-1)^{m/2}\right)q-10K_m(1)^2-45K_m(1)+9K_m(\omega)^2-44}{720},
\end{aligned}\right.
\]
where  $K_m(c)=\sum\limits_{x\in \F_q^*} (-1)^{\Tr_m(x + c/x)}$ is the Kloosterman sum.

\end{theorem}

The remainder of this paper is organized as follows. Section~\ref{sec:prelim} introduces basics
of Niho exponents, Kloomsterman sums and self-conjugate-reciprocal (SCR) polynomials. 
Section~\ref{sec:SCR_poly} presents a general method for  transforming SCR polynomials over $\F_{q^2}$ to new polynomials over $\F_q$ and discusses the relation between their roots.
Section~\ref{sec:4s=5} studies the correlation distribution for the Niho exponent $d$ in \eqref{ourniho} and gives three identities for the numbers of $a$'s for which $P_a(x)$ has exactly $4$, $5$ or $6$ roots in the unit circle $U$. Section~\ref{determiningN4N6} completely determines the correlation distribution by an in-depth investigation of the involved character sums, which are finally expressed in terms of Kloosterman sums $K_m(1)$ and $K_m(\omega)$. Section~\ref{sec:conclusion} summerizes main contributions of this work.

\section{Preliminaries}\label{sec:prelim}

Throughout this paper we assume $q=2^m$ and denote by $\F_q$ and $\F_{q^2}$ the finite fields with $q$ and $q^2$ elements, respectively.
We use $\Tr_m(\cdot)$ for the absolute trace function from $\F_q$
to $\F_2$, and $\Tr_{2m}(\cdot)$ for the absolute trace function from $\F_{q^2}$
to $\F_2$. The \textit{conjugate} of an element $a\in \F_{q^2}$ is defined as $\bar{a} = a^{q}$ and 
the \textit{unit circle} in $\F_{q^2}$ is defined as $U = \{u\in \F_{q^2}: u \bar{u} = 1\}, $
which is a cyclic subgroup in the multiplicative group $\F_{q^2}^*$.

\subsection{Kloosterman sums over finite fields}

For $a,b\in\F_q$, the \textit{Kloosterman sum}
$K_m(a,b)$ over  $\F_q$  is defined by
\begin{equation}\label{eq: kloostermaN_sum}
K_m(a,b)=\sum_{x\in\F_q^*}(-1)^{\Tr_m(ax+b/x)}=\sum_{x\in\F_q^*}\chi\left(ax+b/x\right),
\end{equation} where $\chi$ is the canonical additive character of $\F_q$,
defined by $\chi(x)=(-1)^{\Tr_m(x)}$. 
Kloosterman sums were introduced in 1926~\cite{kloosterman1926} to study a problem
on the representation of integers by quadratic
forms. Later they found applications in coding theory and cryptography
and have been extensively studied in last decades; see, for example,
\cite{Finite,HellZin1999,CharpinGong2008,HollmannXiang2004}. 
For further basic properties of Kloosterman sums over finite fields, we refer the reader to \cite{Finite,zinoviev2019classical}. 

\smallskip

Throughout this paper, we abbreviate $K_m(1, b)$ as $K_m(b)$. 
For $a\ne0$, the substitution $y=ax$ gives
\(
K_m(a,b)=K_m(ab).
\) The substitution $x\mapsto x^{-1}$ also gives $K_m(a,b)=K_m(b,a)$ for  any $a,b\in\F_q$. We also have $K_m(0)=-1$ and $K_m(c^2)=K_m(c)$ for any $c\in \F_q$. 
 Below we recall some identities and recurrence relations
that will be used in subsequent sections.
We start with one basic identity:
\begin{equation}\label{klkeyp1}
    K_m(a,b)=\sum_{x\ne0}\chi(ax+b/x)=
\begin{cases}
q-1,&\text{if }a=b=0,\\
K_m(ab),&\text{otherwise}.
\end{cases}
\end{equation}
The following lemmas give some useful identities of Kloosterman sums over $\F_q$.
\begin{lemma}\cite{HellZin1999}\label{lem: ks_identities1} Let $q=2^m$ and $T_i=\{c\in\F_q:\Tr_m(c)=i\}$, $i=0,1$. The Kloosterman sums $K_m(c)$ over $\F_q$ satisfy the following identities
\begin{enumerate}[label=\textup{(\roman*)}]
\item $\displaystyle \sum_{c\in\F_q} K_m(c)=0$,
\item $\displaystyle \sum_{c\in T_i} K_m(c)
=(-1)^{i+\Tr_m(1)}\,\frac{q}{2}$,
\item $\displaystyle \sum_{c\in\F_q} \chi(c)K_m(c)
=(-1)^{\Tr_m(1)}q$.
\end{enumerate} 
\end{lemma}

\begin{lemma}\label{lem: ks_identities2} Let $q=2^m$ for even $m$, $\sigma=(-1)^{m/2}$ and $\omega$ be a primitive root of $\F_4$, namely, $\omega^2 + \omega + 1 = 0$. Then the Kloosterman sums $K_m(c)$ over $\F_q$ satisfy the following identities:
	\begin{enumerate}[label=\textup{(\roman*)}]
    \item $\sum\limits_{c\in \F_q^*} K_m(1+c+c^{-1}) = K_m(\omega)^2$;
     \item $\sum\limits_{c\in \F_q^*} \chi(c^{-1})K_m(c) = K_m(1)^2-q$;
	\item $\sum\limits_{c\in\F_q^*}\chi(c^{-1})K_m(c+1)=K_m(\omega)^2$;
     \item $\sum\limits_{c\in\F_q^*}K_m(c^3+c)=(2\sigma+1)q+1$.
	\end{enumerate}
\end{lemma}
\begin{proof} These identities can be shown by routine calculations, and the proof is given in Appendix B for completeness. 
\end{proof}
We also need the following classical results about Kloosterman sums over $\mathbb{F}_q$~\cite{Carlitz1969, Carlitz1957,Finite}.
\begin{lemma}\cite[Sec.~6-7]{Carlitz1969}\label{ks_recurrence}
	Let $q = 2^m$, $k_m = K_m(1)$, and  for even $m$, let $\ell_m = K_m(\omega)$, where $\omega \in \mathbb{F}_q$ satisfies $\omega^2 + \omega + 1 = 0$. Then we have the following recurrences:
	\[
    \begin{aligned}
        & k_1 = 1,\quad k_2 = 3,\quad k_{j+2} = -k_{j+1} -2k_j,\quad j \geq 1; \\
        & \ell_2 = -1, \quad \ell_4 = 7, \quad \ell_{2(j+2)} = \ell_{2j}  - 4 \ell_{2(j+1)}, \quad j\geq 1.
    \end{aligned}
	\]
\end{lemma}

\begin{lemma}\cite{Carlitz1957}\label{cor:Kbound} 
For every $c\in \F_q$, one has $|K_m(c)|\le2\sqrt q$.
\end{lemma}

\subsection{Niho exponents and their associated key polynomials}
Below we recall the well-known result on Niho exponents over finite fields of characteristic two, which will be referred to as Niho's theorem~\cite{Niho,Rosendahl2006}. 

\begin{lemma}\cite{Niho,Rosendahl2006}\label{lem:niho} 
	For a Niho exponent $d=s(q-1)+1$ with $s$ modulo $q+1$,
	the Walsh transform of the function $\Tr_{2m}(x^d)$ at a point $a\in \F_{q^2}$ satisfies
		\begin{equation}\label{eq:niho-sum}
		W_d(a):=\sum_{x\in\F_{q^2}}
		(-1)^{\Tr_{2m}(x^d+ax)}
		=q\bigl(N_s(a)-1\bigr).
	\end{equation} where $N_s(a)$ is the number of distinct roots in $U$ of the  polynomial 
	\begin{equation}\label{eq:niho-general}
		f_{s,a}(x) = x^{2s-1}+xz^s+\bar a x^{s-1}+1.
	\end{equation}
\end{lemma}
The polynomial $f_{s,a}(x)$ will be termed the \textit{key polynomial} associated with 
a normalized Niho exponent $d=s(q-1)+1$. By Niho's theorem, determining the value distribution of the Walsh transform of the Niho-type function $\Tr_{2m}(x^d)$ is equivalent to determining the possible numbers of roots of $f_{s,a}(x)$ in $U$ and the frequency of each number as $a$ runs through $\F_{q^2}^*$.
Notice that the polynomial $f_{s,a}(x)$ is an instance of self-conjugate-reciprocal polynomials over $\F_{q^2}$~\cite{HKL2021}. Below we recall some basics of self-conjugate-reciprocal polynomials.

\subsection{Self-conjugate-reciprocal polynomials over finite fields}

The \textit{reciprocal polynomial} of a polynomial $f(x) = \sum_{i=0}^d f_i x^i$ in $\F_{q^2}[x]$
is defined as $f^*(x) := x^d f(x^{-1})$ and the \textit{conjugate polynomial}  of $f(x)$
is defined as $\overline{f}(x):= \sum_{i=0}^d \overline{f_i} x^i$, where $\overline{f_i}$
is the conjugate of $f_i$ in $\F_{q^2}$. The \textit{conjugate-reciprocal polynomial} of $f(x)$ is then defined as 
\[
 f^{\dagger}(x) := x^d \sum_{i=0}^d \overline{f_i} x^{-i} = \sum_{i=0}^d \overline{f_i} x^{d-i} = \sum_{i=0}^d \overline{f_{d-i}} x^{i}.
\]

\begin{definition}\cite{HKL2021}
    A polynomial $f(x) = \sum_{i=0}^d f_i x^i$ in $\F_{q^2}[x]$ is  \textit{self-conjugate-reciprocal} (SCR) if $f(x)=f^{\dagger}(x)$, namely, 
    its coefficients satisfy $f_i=\overline{f_{d-i}}$ for $0\leq i\leq d$. In particular, a monic SCR polynomial necessarily satisfies 
    $f_d = f_0 = 1$, and will be termed a \textit{normalized SCR}  polynomial.
\end{definition}
For instance, normalized SCR polynomials
of degrees two, three, and four have the following forms, respectively:
\[
\begin{aligned}
&x^2+\alpha x+1,
&&\alpha\in\F_{q},\\
&x^3+\alpha x^2+\bar\alpha x+1,
&&\alpha\in\F_{q^2},\\
&x^4+\alpha x^3+\beta x^2+\bar\alpha x+1,
&&\alpha\in\F_{q^2},\ \beta\in\F_{q}.
\end{aligned}
\]
Note that SCR polynomials have some interesting properties. For an SCR polynomial $f(x)$ and an element $\lambda\in \F_{q^2}^*$,  it is readily seen that $\lambda f(x)$ is SCR if and only if $\lambda\in \F_{q}^*$. 
The operators of \textit{reciprocal} and \textit{conjugate} on polynomials in $\F_{q^2}[x]$ are involutions.
The SCR property is preserved by polynomial multiplications. Namely,
if $f(x)$ and $h(x)$ are SCR polynomials, 
then $f(x)h(x)$ is an SCR polynomial, which is due to the
following equalities:
\[
 (fh)^*(x) = f^*(x)h^*(x),\,\,
 \overline{f h}(x) = \overline{f}(x)\, \overline{h}(x),\,\,
 \overline{(f^{*})}(x) = (\overline{f})^*(x),
\] where $fh$ denotes the multiplication of $f(x)$ and $g(x)$. If $f(x)\in\F_{q^2}[x]$ is an SCR polynomial, then the following
properties hold:
\begin{enumerate}[label=\textup{(\roman*)}]
\item $f^*(x)=\overline f(x)$, and both $f^*(x)$ and $\overline f(x)$
are SCR polynomials.

\item $f^*(x)=f(x)$ if and only if $\overline f(x)=f(x)$,
or equivalently, $f(x)\in\F_q[x]$.
\end{enumerate}
For a given SCR polynomial, its roots in the algebraic closure has certain interesting structure.
\begin{lemma}\cite{HKL2021}\label{lem:SCR_alg_structure}
   Let $f(x)\in\F_{q^2}[x]$ be an SCR polynomial of degree $d
\geq 2$ and let
$R$ be the set of all its roots in $\overline{\F_{q^2}}$. Let $\Pi$ be the map
from $\overline{\F_{q^2}}^{\,*}$ to itself defined by $\Pi(a)=a^{-q}$.
For $a\in\overline{\F_{q^2}}^{\,*}$, let $O_{\Pi}(a)=\{\Pi^i(a): i = 0,1,\dots, k-1\}$
be the orbit of $a$ under $\Pi$, where $k$ is the least positive
integer such that $\Pi^k(a)=a$. Then we have the following statements:
    \begin{enumerate}[label=(\roman*)]
        \item An element $a\in \overline{\F_{q^2}}$ is a root of $f(x)$ if only if $\Pi(a)$ is a root of $f(x)$;
        \item The root set $\mathcal{R}$ of $f(x)$ can be represented as a disjoint union of 
        distinct orbits $O_{\Pi}(a)$ with $a\in\mathcal{R}$, i.e., $\mathcal{R}=O_{\Pi}(a_1)\cup \cdots \cup O_{\Pi}(a_t) $, where $a_1,\dots,a_t$ are representatives of distinct orbits;
        \item The degree $d$ of $f(x)$ and the number $t$ of distinct orbits in its root set $\mathcal{R}$ satisfy $\binom{d+1}{2} + t \equiv \Tr_m(R) \pmod{2}$, where 
        $R=\sum\limits_{u\neq v\in \mathcal{R}}\frac{uv}{(u+v)^2}.$    
    \end{enumerate}
\end{lemma}
Under an action $\Pi$ defined above, it is clear that the orbit $O_{\Pi}(a)$ has size one 
for $a\in U$ and has size two $a\in \F_{q^2}^*\setminus U$. As a result, Lemma~\ref{lem:SCR_alg_structure} can be used to characterize the possible number of roots of $f(x)$ in $U$ for some cases, such as $f(x) = x^7 + ax^4 + \bar{a}x^3 + 1$ for the Niho decimation $d=4q-3$~\cite{HKL2021}. 

\smallskip

 Recall that a polynomial $f(x)\in\F_{q^2}[x]$ is said to split in a subset of $\overline{\F_{q^2}}$ if
all the roots of $f(x)$ belong to the set, where repeated roots are allowed. 
Here we are interested in the cases where polynomials in $\F_{q^2}[x]$ split in the unit circle $U$.
Suppose that a monic polynomial $f(x)=x^d+\sum_{i=0}^{d-1}f_i x^i\in\F_{q^2}[x]$ 
splits in $U$, i.e., 
$f(x)=\prod_{i=1}^d(x+u_i)$ with $u_i$'s in $U$. 
Since $\overline{u_i}=u_i^{-1}$, its conjugate-reciprocal polynomial satisfies 
\[
\begin{aligned}
f^{\dagger}(x)
=x^d\prod_{i=1}^d(x^{-1}+u_i^{-1})
=\left(\prod_{i=1}^d u_i^{-1}\right)
  \prod_{i=1}^d(x+u_i)
=f_0^{-1}f(x).
\end{aligned}
\]
Thus, if $f_0=1$, equivalently $\prod_{i=1}^d u_i=1$, then $f(x)$ is a normalized SCR polynomial.  More generally, suppose that there exists an element $\mu\in U$ such that $\mu^d=f_0$ (such an element always exists when $\gcd(d,q+1)=1$). Then, one can transform $ f(x)$ to a normalized SCR polynomial as follows: 
$$f(x) \mapsto \mu^{-d}f(\mu x ) = x^d + \sum_{1\leq i <d} \mu^{i-d} f_ix^i + 1.$$
Thus we have the following result.
\begin{lemma}\label{standfornscr} 
Suppose a monic polynomial $f(x)=x^d+f_{d-1}x^{d-1}+\cdots + f_1x+f_0\in\F_{q^2}[x]$ splits in $U$ and there exists an element $\mu\in U $ satisfying $\mu^d=f(0)$. 
Then $\mu^{-d}f(\mu x)$ is a normalized SCR polynomial that splits in $U$.
\end{lemma}
A natural question arises here: when does a normalized SCR polynomial split in $U$? The next section is devoted to the study of low-degree normalized SCR polynomials that split in $U$. The resulting criteria will be used to analyze the hexic polynomial $P_a(x)$ derived from the Niho exponent $d$ in \eqref{ourniho}.

\section{Low-degree self-conjugate-reciprocal polynomials splitting in the unit circle}\label{sec:SCR_poly}
This section characterizes the condition under which a low-degree SCR polynomial $f(x)$ in $\F_{q^2}[x]$ splits in the unit circle $U$ of $\F_{q^2}$. We will frequently use the notation 
\begin{equation}\label{defH1}
    H_1 = \{w \in \F_q^*\, : \, \Tr_m(1/w) = 1\}.
\end{equation}
It was pointed out~\cite[Prop.~7]{Dobbertin} that 
$H_1 = \{u+\bar{u}: u\in U \setminus\{1\}\}$, which can easily verified by considering the roots of the quadratic polynomial $X^2+wX+1$ over $\F_q$.

Below we start with a method of transforming a normalized SCR polynomial in $\F_{q^2}[x]$ to a polynomial in $\F_{q}[x]$ and the relation between their roots. 

\begin{proposition}\label{prop:trace-transform}
Let $f(x)=\sum_{i=0}^d f_ix^i\in \F_{q^2}[x]$ be a normalized SCR polynomial of degree $d \geq 2$ and $D_k(y)$ denote the first-kind Dickson polynomial over $\F_2$ of degree \(k\). Set
\begin{equation}\label{defsk}
 s_k=\sum_{i=k}^d f_i f_{i-k}, \quad 0\leq k\leq d. 
\end{equation} 
Then each $s_k$ lies in $\F_q$ and the polynomial 
\begin{equation}\label{eq-trans1}
g(y):=s_0+\sum_{k=1}^d s_kD_k(y),
\end{equation}
satisfies $x^d g(x+x^{-1})=f(x)\overline f(x).$
Moreover, the following properties hold.
\begin{enumerate}[label=\textup{(\roman*)}]
\item If $x_1,\ldots,x_d$ are the roots of $f(x)$ in $\overline{\F_{q^2}}$, counted with multiplicity, then
\[
 g(y)=\prod_{i=1}^d\bigl(y+x_i+x_i^{-1}\bigr).
\]
\item For $u\in U$ and $t\in\F_q$, let $m_f(u)$ and $m_g(t)$ denote the multiplicities of $u$ and $t$ as roots of $f(x)$ and $g(y)$, respectively, with multiplicity zero if the corresponding element is not a root.
Then the map $u\mapsto u+u^{-1}=t$ gives the following
correspondence between the roots of $f(x)$ in $U$ and those
of $g(y)$ in $\{0\}\cup H_1$:
\begin{enumerate}[label=\textup{(\alph*)}]
\item $f(1)=0$ if and only if $g(0)=0$; moreover, $m_g(0)=m_f(1).$

\item Each root $u\in U\setminus\{1\}$ of $f(x)$ gives a root
$t=u+u^{-1}\in H_1$ of $g(y)$.
Conversely, for each root $t\in H_1$ of $g$, let
$u,u^{-1}\in U\setminus\{1\}$ be the two distinct roots
of $X^2+tX+1$; then $u,u^{-1}$
are roots of $f(x)$, and
$m_g(t)=m_f(u)+m_f(u^{-1}).$
\end{enumerate}
\end{enumerate}
\end{proposition}

\begin{proof}
By the SCR condition of $f(x)$, namely, $f_i=\overline{f_{d-i}}$ for $0\leq i\leq d$, 
one has 
\[
\overline{s_k}
=\sum_{i=k}^d\overline{f_i}\,\overline{f_{i-k}}=\sum_{i=k}^d f_{d-i}f_{d-i+k}
=\sum_{j=0}^{d-k}f_jf_{j+k}
=s_k,
\]
implying that $s_k\in\F_q$.  Note that the first-kind Dickson polynomial 
satisfy $ D_k(x+x^{-1})=x^k+x^{-k}$ for $k\geq 0$ \cite{Dobbertin,Bluher2021}. Thus, 
$$
x^dg(x+x^{-1}) = x^d\left(s_0 +\sum_{k=1}^d s_k\left(x^k+x^{-k}\right)\right).
$$
On the other hand, for the normalized SCR polynomial $f(x)$, one has
$$
\overline{f}(x)=f^*(x)=x^d f(1/x),
$$
and then
$$
f(x)\overline{f}(x) = x^d f(x) f(x^{-1}) = x^d(s_0 + \sum_{k=1}^d s_k(x^k+x^{-k}) = x^d g(x+x^{-1}).
$$

(\textup{i}) Let $x_1,\ldots,x_d$ are the roots of $f(x)$ in an algebraic
closure of $\F_{q^2}$, counted with multiplicity. Then $x_i$, $1\le i\le d$, are
nonzero and satisfy $x_1\cdots x_d=1$, and $1/x_i$ , $1\le i\le d$, are root of $f^*$. Consequently,
\[
 \overline f(x)=x^d f(x^{-1})
 =\prod_{i=1}^d(x+x_i^{-1}).
\]
It follows that
\begin{align*}
 f(x)\overline f(x)
 &=\prod_{i=1}^d(x+x_i)(x+x_i^{-1})\\
 &=\prod_{i=1}^d\bigl(x^2+(x_i+x_i^{-1})x+1\bigr)\\
 &=x^d\prod_{i=1}^d\bigl(x+x^{-1}+x_i+x_i^{-1}\bigr).
\end{align*}
Thus, over the algebraic closure of $\F_{q^2}$, we have
\[
 g(y)=\prod_{i=1}^d\bigl(y+x_i+x_i^{-1}\bigr).
\]

\textup{(ii)} Suppose first that $f(x)$ has a root $x_i\in U$.
Then $x_i^q=x_i^{-1}$, so $t_i=x_i+x_i^{-1}\in\F_q$.
If $x_i=1$, then $t_i=0$. If $x_i\in U\setminus\{1\}$, then
$t_i=x_i+x_i^{-1}\in\F_q^*$.
Using $x_i^q=x_i^{-1}$, we obtain
\[
\begin{aligned}
\Tr_m(t_i^{-1})
&=\Tr_m\!\left(\frac{x_i}{(1+x_i)^2}\right)\\
&=\sum_{j=0}^{m-1}
  \left(
    \frac{1}{1+x_i}+\frac{1}{(1+x_i)^2}
  \right)^{2^j}\\
&=\frac{1}{1+x_i}+\frac{1}{(1+x_i)^q}\\
&=\frac{1}{1+x_i}+\frac{1}{1+x_i^{-1}}
 =1.
\end{aligned}
\] Thus, $t_i\in H_1$ and $g$ has a root in $H_1$.

Conversely, suppose that $g(y)$ has a root $t_i\in\{0\}\cup H_1\subset \F_q$. Part \textup{(i)} shows that in the algebraic closure of $\F_{q^2}$ there exists a root $x_i$ of $f(x)$ such that 
$t_i=x_i+x_i^{-1}$.  If $t_i=0$, then $x_i=1\in U$.
If $t_i\ne0$, the trace condition $\Tr_m(t_i^{-1})=1$ implies that
$X^2+t_iX+1$ has two distinct roots $u,u^{-1}\in U\setminus\{1\}$.
Since $x_i$ satisfies this quadratic equation, we have
$x_i\in\{u,u^{-1}\}\subset U\setminus\{1\}$.
Thus, in this case $f(x)$ has a root in $U$, which is different from $1$.   

The multiplicity statement follows from the fact that the correspondence $t=u+u^{-1}$ preserves multiplicities between the roots $u\in U$ of $f(x)$ and the corresponding roots $t\in H_1 \cup \{0\}$ of $g(y)$. This proves \textup{(ii)}. 
\end{proof}

Proposition~\ref{prop:trace-transform}
provides a handy tool to transform the roots of a normalized SCR polynomial $f(x)$ in $U$ 
to that of $g(y)$ in $\{0\}\cup H_1$, which 
was well studied for low-degree polynomials. 
To compute $g(y)$, we need to use the Dickson polynomials $D_k(y)$ of the first kind with parameter $1$. In characteristic two, they satisfy the recurrence\cite{Dobbertin,Bluher2021}  
\begin{equation}\label{dicksondef}
D_0(y)=0,\quad D_1(y)=y,\quad
D_k(y)=yD_{k-1}(y)+D_{k-2}(y)\quad(k\ge2).
\end{equation}

By Proposition~\ref{prop:trace-transform}, we immediately have the following criterion for a quadratic polynomial to split in $U$.
\begin{corollary}\label{cor:deg2}
A quadratic normalized SCR polynomial
$f(x)=x^2+\alpha x+1\in\F_{q^2}[x]$ splits in $U$ if and only if
either $\alpha=0$ or $\alpha\in H_1$.
In the latter case, its roots are two distinct elements
$u,u^{-1}\in U\setminus\{1\}$. More generally, a quadratic polynomial
$f(x)=x^2+\alpha x+\beta\in\F_{q^2}[x]$ with $\alpha\ne0$
has two distinct roots in $U$ if and only if
\(
\beta=\frac{\alpha}{\bar\alpha}
\text{ and }
\alpha\bar\alpha\in H_1.
\)
Both roots belong to $U\setminus\{1\}$ if and only if,
in addition, $1+\alpha+\beta\ne0$.
\end{corollary}
\begin{proof}
    For a normalized SCR polynomial $f(x) = x^2+\alpha x + 1$, it is clear that $\alpha \in \F_q$, and the associated polynomial of $f(x)$ is $g(y)=y^2+\alpha^2$, which is computed by \eqref{defsk}, \eqref{eq-trans1} and \eqref{dicksondef}. The root of $g(y)$ is $\alpha$ with multiplicity two. By Proposition~\ref{prop:trace-transform}, we get the desired result.   
    
    Now consider a general quadratic polynomial
$f(x)=x^2+\alpha x+\beta\in\F_{q^2}[x]$ with $\alpha\ne0$.  By Lemma \ref{standfornscr}, $f(x)$ has two distinct roots in $U$ if and only if  $\beta\in U$ and $\mu^{-2}f(\mu x)=x^2+\frac{\alpha}{\mu}x+1$ is a normalized SCR polynomial with two distinct roots in $U$, where \(\mu\) is the unique element of \(U\) satisfying $\mu^2=\beta$. This holds if and only if $\frac{\alpha}{\mu}\in H_1$, whihc implies that $\left(\frac{\alpha}{\mu}\right)^q=\frac{\alpha}{\mu}$, i.e., $\mu^2=\beta=\frac{\alpha}{\bar \alpha}$.  In addition,  $\frac{\alpha}{\mu}\in H_1$ if and only if $\frac{\alpha^2}{\mu^2}=\alpha \bar \alpha\in H_1$. Since $f(1)=1+\alpha+\beta$, neither of the two distinct roots in $U$ equals $1$ if and only if \(1+\alpha+\beta\ne0\). This completes the proof.
\end{proof}

The following corollary characterizes when a cubic normalized SCR polynomial splits in $U$.
\begin{corollary}\label{cor:deg3}
    Let 
    $f(x)=x^3+\alpha x^2 + \bar{\alpha} x + 1$ in $\F_{q^2}[x]$ with $\alpha\neq 0$. 
    Then $f(x)$ splits in $U$ if and only if 
    the cubic polynomial 
        \(
         g(y) = y^3 + (\alpha+\bar \alpha) y^2 + (\alpha+1)(\bar \alpha +1) y + (\alpha+\bar \alpha)^2 \in \F_q[y]
        \) splits in $\{0\}\cup H_1$. Furthermore, $f(x)$ has exactly three distinct roots in $U$ if and only if one of the following holds: 
        \begin{enumerate}[label=\textup{(\roman*)}]
        \item $\alpha \in \F_q^*$ and $\alpha+1 \in H_1$. In this case,
     $ f(x)=(x+1)(x+u)(x+u^{-1})$
        for some $u\in U\setminus\{1\}$ satisfying $\alpha+1=u+u^{-1}$.
        \item $\alpha \not\in \F_q\cup U$ and $g(y)$ has three distinct roots in $H_1$.
        \end{enumerate}
\end{corollary}

\begin{proof} For $k=0,1,2,3$, we calculate $s_k$ and $D_k(y)$ by \eqref{defsk} and \eqref{dicksondef} as follows: 
$s_0=\left(\alpha+\overline{\alpha}\right)^2$, $s_1=\overline{\alpha}+\alpha\overline{\alpha}+\alpha$, $s_2=\alpha+\overline{\alpha}$, $s_3=1$; $D_1(y)=y$, $D_2(y)=y^2$, $D_3(y)=y^3+y$. Then, 
 \begin{equation}\label{gyfor3}
    g(y) =s_0+\sum\limits_{k=1}^3s_kD_k(y)= y^3 + T y^2 + (T+N+1) y + T^2,
\end{equation}
  where $T= \alpha +\bar{\alpha}$ and $  N=\alpha \bar{\alpha}.$
  It follows from Proposition~\ref{prop:trace-transform} that 
  $f(x)$ splits in $U$ if and only if $g(y)$ splits in $\{0\}\cup H_1$.

Since $f'(x) = x^2+\bar{\alpha}$ and $f(\bar{\alpha}^{\frac{1}{2}}) = \alpha \bar{\alpha} + 1 = N+1$, $f(x)$ has no repeated roots if and only if  $N+1 \neq 0$, or equivalently $\alpha\notin U$.  If $\alpha \in \F_q^*$, then $f(x) = (x+1)(x^2+(\alpha + 1)x+1)$. By Corollary \ref{cor:deg2}, $f(x)$ has three distinct roots in $U$
if and only if $\alpha+1\in H_1$. This gives case~\textup{(i)}. 

Suppose now that $\alpha\notin\F_q$, so $T\ne0$. Let $g(y)$ be the polynomial given in \eqref{gyfor3}. Since $g(0)=T^2\ne0$, none of the roots of $g(y)$ is zero. Let $w_1,w_2,w_3$ be the roots of $g(y)$ in an algebraic
closure, counted with multiplicity.  Then,  
  $$\prod\limits_{1\le i<j \le 3}(w_i+w_j) =w_1w_2w_3+\left(\sum\limits_{i=1}^3w_i\right)\left(\sum\limits_{1\le i<j\le 3}w_iw_j\right)= T(N+1).$$  If $f(x)$ has three distinct roots in $U$, then $N\ne1$.
Hence $g(y)$ has three distinct nonzero roots, all of which
lie in $H_1$ by the splitting criterion in Proposition~\ref{prop:trace-transform}. This proves the conditions in \textup{(ii)}.
 Conversely, if \textup{(ii)} holds, then $g(y)$ has all its
roots in $H_1$, so $f(x)$ splits in $U$.
Moreover, $\alpha\notin U$ implies $N\ne1$, so $f(x)$
has no repeated roots. It therefore has three distinct
roots in $U$.
  \end{proof}
  
Given a normalized SCR polynomial $f(x)\in\F_{q^2}[x]$ of higher degree, we can similarly construct its associated polynomial $g(y)$ in \eqref{eq-trans1} and apply Proposition~\ref{prop:trace-transform} to determine whether $f(x)$ splits in $U$. For example, for the quartic normalized SCR polynomial $f(x) = x^4+\alpha x^3+\beta x^2+\bar{\alpha}\,x+1$, where $\beta\in \F_q$,  by Proposition~\ref{prop:trace-transform}, 
the associated polynomial $g(y)$ is given by 
\begin{equation*}\label{associatedpoly4}
   g(y)=y^4+T\,y^3+N\,y^2+T\beta\,y+T^2+\beta^2,\quad  \text{ with } T= \alpha+\bar{\alpha},\quad N = \alpha \bar{\alpha}. 
\end{equation*}
Consequently, $f(x)$ splits in $U$ if and only if $g(y)$ splits in $\F_q$ and all its nonzero roots lie in $H_1$. However, determining whether a quartic polynomial splits in $\{0\}\cup H_1$ is nontrivial. We therefore derive an alternative splitting criterion for the quartic case, which will be applied to the key polynomial associated with the Niho exponent in \eqref{ourniho}.

\begin{proposition}\label{prop:resolvent}
Let $f(x)=x^4+\alpha x^3+\beta x^2+\bar{\alpha}\,x+1 \in \F_{q^2}[x]$ be a normalized SCR polynomial, where $\beta \in \F_q$, and let
\[
 h(y)=y^3+\beta y^2+\alpha\bar{\alpha}y
      +(\alpha+\bar{\alpha})^2.
\] 
Then the following statements hold: 
\begin{enumerate}[label=\textup{(\roman*)}]
\item $f(x)$ splits in $U$ if and only if all roots of $h(y)$ lie in $\{0\}\cup H_1$.
\item $f(x)$ has no repeated roots if and only if $h(y)$ has
no repeated roots. These equivalent conditions hold precisely
when
 $\beta\alpha\bar{\alpha}+(\alpha+\bar{\alpha})^2\ne0$.
\end{enumerate}
Consequently, $f(x)$ has four distinct roots in $U$ if and
only if $h(y)$ has three distinct roots in $\{0\}\cup H_1$. Moreover,  $h(y)$ has no zero root if and only if $\alpha\notin\F_q$. 
\end{proposition}

\begin{proof} Let $u_1,u_2,u_3,u_4$ be the roots of $f(x)$ in an algebraic closure,
counted with multiplicity. It is clear that 
\[\begin{aligned}
    \sum_{1\leq i \leq 4} u_i=\alpha, \quad
    \sum_{1\leq i<j \leq 4} u_iu_j=\beta,\quad
    \sum_{1\leq i<j<k\leq 4} u_iu_ju_k =\bar{\alpha} \quad\text{and}\quad
    u_1u_2u_3u_4=1.
\end{aligned}\]
Proposition~\ref{prop:trace-transform} associates $f(x)$ with
\[
g(y)=\prod_{i=1}^4(y+u_i+u_i^{-1})
=y^4+Ty^3+Ny^2+T\beta y+T^2+\beta^2,
\]
where $T= \alpha+\bar{\alpha}$ and $N = \alpha \bar{\alpha}$. 
It follows that $f(x)$ splits in $U$ if and only if $g(y)$ splits in $\{0\}\cup H_1$. In particular, each root $u_i$ of $f(x)$ gives a root
$u_i+u_i^{-1}$ of $g(y)$.

Since $u_1u_2u_3u_4=1$, the six pairwise products form three
reciprocal pairs. Let
\[
v_1=u_1u_2,~v_2=u_1u_3 ~\text{and}~v_3=u_1u_4. 
 \]
The other three pairwise products are $v_1^{-1},v_2^{-1},v_3^{-1}$. 
Let $w_k=v_k+v_k^{-1}$ with $k=1,2,3$ and define 
\[h(y)=\prod_{k=1}^3(y+w_k).\]
Then, we have
\[
\begin{aligned}
 w_1+w_2+w_3 &=\sum_{1\leq i<j \leq 4} u_iu_j=\beta, \\
 \sum_{1\leq i<j\leq 3}w_iw_j&=\left(\sum\limits_{i=1}^4u_i\right)\left(\sum\limits_{1\le j<j<k\le 4}u_iu_ju_k\right)
 =\alpha\bar{\alpha}=N, \\
 w_1w_2w_3 
 &= \left(\prod\limits_{i=1}^4u_i\right)\left(\sum\limits_{i=1}^4u_i^2\right)+\left(\sum\limits_{1\le j<j<k\le 4}u_iu_ju_k\right)^2=\alpha^2+\bar{\alpha}^{2}=T^2.
 \end{aligned}
\]
Thus, $h(y)=y^3+\beta y^2+Ny+T^2\in \F_q[x]$. 

Define
\[
F(z)=\prod_{1\le i<j\le4}(z+u_iu_j).
\]
Then
\[
\begin{aligned}
F(z)
&=\prod_{k=1}^3(z+v_k)(z+v_k^{-1})\\
&=\prod_{k=1}^3(z^2+w_kz+1)\\
&=z^3h(z+z^{-1})\\
&=z^6+\beta z^5+(N+1)z^4+T^2z^3
  +(N+1)z^2+\beta z+1.
\end{aligned}
\]
Thus $F(z)\in\F_q[z]$ is a normalized SCR polynomial. By Proposition~\ref{prop:trace-transform}, the polynomial associated with $F(x)$ is
\[
\begin{aligned}
G(y)
&=\prod_{k=1}^3
 (y+v_k+v_k^{-1})(y+v_k^{-1}+v_k)\\
&=\prod_{k=1}^3(y+w_k)^2
=h(y)^2.
\end{aligned}
\]
Consequently, $F(z)$ splits in $U$ if and only if $h(y)$ splits in 
$\{0\}\cup H_1$.

If $f(x)$ splits in $U$, then all pairwise products of its roots
belong to $U$, so $F(z)$ splits in $U$.
Conversely, if $F(z)$ splits in $U$, then $u_iu_j\in U$ for every
$1\le i<j\le 4$. For each $i\in\{1,2,3,4\}$,
\[
u_i^2=\prod_{\substack{1\le j\le4\\ j\ne i}}(u_iu_j)\in U.
\]
Hence $(u_i^{q+1})^2=1$. This implies
$u_i^{q+1}=1$, so every root of $f(x)$ lies in $U$. This shows that $f(x)$ splits in $U$ if and only if $F(z)$ splits in $U$.

Combining the above arguments, we obtain that
\[
\begin{aligned}
g(y)\text{ splits in }\{0\}\cup H_1
&\Longleftrightarrow f(x)\text{ splits in }U\\
&\Longleftrightarrow F(z)\text{ splits in }U\\
&\Longleftrightarrow h(y)\text{ splits in }\{0\}\cup H_1.
\end{aligned}
\]
This proves the first assertion (\textup{i}) using Proposition~\ref{prop:trace-transform}.

For {(\textup{ii})}, by the definitions of the $w_k$, we have
\[
\prod_{1\leq i<j\leq 4}(u_i+u_j) = \prod_{1\leq i<j\leq 3}(w_i+w_j) 
= \beta \alpha\overline{\alpha} + (\alpha+\overline{\alpha})^2.
\]
Hence $f(x)$ has no repeated roots if and only if
$h(y)$ has no repeated roots. These equivalent
conditions hold precisely when $\beta \alpha\overline{\alpha} + (\alpha+\overline{\alpha})^2\neq 0$.  This proves \textup{(ii)}.

The assertion about four distinct roots in $U$ follows
from \textup{(i)} and \textup{(ii)}.  Finally, $h(0)=0$ if and only if $\alpha=\bar{\alpha}$, equivalently $\alpha\in\F_q$.
\end{proof}

\begin{remark}
Using Proposition~\ref{prop:trace-transform}, we derive
a simpler splitting criterion in
Proposition~\ref{prop:resolvent} for the quartic normalized SCR polynomial $f(x)$: all four roots of $f(x)$ lie in $U$ if and only if
all three roots of $h(y)$ lie in $\{0\}\cup H_1$. In fact, $h(y)$ is the classical cubic resolvent of $f(x)$ \cite[Sec.~13.1]{Cox2012}, whose roots are obtained from the three pairings of the four roots of $f(x)$.
\end{remark}

\begin{remark}
For a Niho decimation $d=s(q-1)+1$ with $q=2^m$,
the associated key polynomial $f_{s,a}(x)=x^{2s-1}+ax^s+\bar a x^{s-1}+1
$ in Lemma \ref{lem:niho} is a normalized SCR polynomial. Recall that $N_s(a)=\bigl|\{x\in U:f_{s,a}(x)=0\}\bigr|$.
For small values of $s$, the preceding splitting criteria can be used
to study the relationship between $a$ and $N_s(a)$.
We illustrate this approach below for $s=2$ and $s=3$.
     In the next section, we will consider the case $s\equiv \frac{5}{4}\pmod{q+1}$. 
\end{remark}

When $s=2$, the polynomial $f_{s,a}(x)$ becomes a cubic polynomial $f_a(x) = x^{3} + a x^{2} + \bar{a} x + 1$. For $a\ne0$, its splitting criterion in $U$ has been discussed in Corollary~~\ref{cor:deg3}. We consider the following five cases, including $a=0$:
\begin{enumerate}[label=(\roman*)]\setlength{\itemsep}{1pt}
\item $a = 0$.  Then, $f(x) = (x+1)(x+\omega)(x+\omega^2)$, where $\omega$ is a primitive element of $\F_4$.
Since $\omega\in U$ if and only if $m$ is odd, we have
\[
N_2(0)=
\begin{cases}
3,&m\text{ is odd},\\
1,&m\text{ is even}.
\end{cases}
\]

\item $a =1$. Then $f_1(x)=(x+1)^3$, so $N_2(1)=1$.

\item $a \in U\setminus \{1\}$. Then $
f_a(x)=(x+a)\bigl(x+\sqrt{\bar a}\bigr)^2$. 
Both $a$ and $\sqrt{\bar a}$ lie in $U$, and they coincide
if and only if $a^3=1$. Therefore,
\[
N_2(a)=
\begin{cases}
1,&a^3=1,\\
2,&a^3\ne1.
\end{cases}
\]

\item $a \in \F_q^*\setminus\{1\}$. By Corollary~\ref{cor:deg3},
\[
N_2(a)=
\begin{cases}
3,&a+1\in H_1,\\
1,&a+1\notin H_1.
\end{cases}
\]
When $a+1 \in H_1$, we have $f_a(x)=(x+1)(x+u)(x+u^{-1})$ with $u\in U\setminus \{1\}$; 

\item $a \in \F_{q^2}\setminus (\F_q\cup U)$. By Corollary~\ref{cor:deg3},
$f(x)$ has three distinct roots in $U\setminus \{1\}$ if and only if
$g(y)=y^3+Ty^2+(N+T+1)y+T^2$ has three distinct roots in $H_1$, where $T=\alpha+\bar{\alpha}$ and $N=\alpha \bar{\alpha}$. 

\end{enumerate} 
\smallskip 

When $s=3$, the key polynomial becomes the quintic
$f_a(x)=x^5+ax^3+\bar a x^2+1$.
Using the splitting criteria for quadratic and cubic normalized
SCR polynomials in Corollaries~\ref{cor:deg2} and~\ref{cor:deg3},
we obtain the following explicit characterization of the parameters
$a$ for which $f_a(x)$ has exactly three or five distinct roots
in $U$ when $m$ is even.

\begin{proposition}\label{prop:deg-5 s=3}
Let $q=2^m$, where $m$ is even, and let $f_a(x)=x^5+ax^3+\bar a x^2+1$ with $a\in \F_{q^2}$ and $H_1=\{
w \in \F_q^*\,:\, \Tr_m(1/w)=1\}$. Define
\[
\begin{aligned}
\mathcal A={}&\left\{\alpha\in\F_q^*: \alpha+1\in H_1\right\}\\
&{}\bigcup\left\{\alpha\in\F_{q^2}\setminus(\F_q\cup U) : \begin{array}{c}
     g(y)=y^3 + (\alpha+\bar{\alpha})y^2 + (\alpha+1)(\bar{\alpha}+1)y + (\alpha+\bar{\alpha})^2  \\
     \text{has three distinct roots in 
} H_1 
\end{array}\right\}.
\end{aligned}
\] 
For $\alpha\in \mathcal{A}$, let 
$
h_{\alpha}(x) = x^3 + \alpha x^2 + \bar{\alpha} x + 1.
$ 
Then $f_a(x)$ has at least three distinct roots in $U$ if and only if there exist $\alpha\in\mathcal A$ and $\mu\in U$
such that \[
\mu^5={\bar\alpha \over \alpha}\quad \text{and}\quad a = \frac{(\bar{\alpha}^2+ \alpha^2\bar{\alpha} + \alpha)\mu^2}{\bar{\alpha}}.
\]
For any such pair $(\alpha,\mu)$, we have
\[
\mu^{-5}f_a(\mu x)=h_{\alpha}(x)\left(x^2+\alpha x + \frac{\alpha}{\bar{\alpha}}\right). \]  
Moreover, the following hold:
\begin{enumerate}[label=\textup{(\roman*)}]
    \item $f_a(x)$ has exactly three roots in $U$
    if  $\Tr_m(\frac{1}{\alpha \bar{\alpha}}) = 0$;
    \item $f_a(x)$ has exactly five roots in $U$
    if  $\Tr_m(\frac{1}{\alpha \bar{\alpha}}) = 1$.
\end{enumerate}
\end{proposition}

\begin{proof}
Suppose first that $\alpha\in\mathcal A$ and $\mu\in U$
satisfy the stated relations.
A direct expansion gives the stated factorization.
By Corollary~\ref{cor:deg3} and the definition of $\mathcal A$,
$h_\alpha(x)$ has three distinct roots in $U$.
Multiplying these roots by $\mu$ gives three distinct roots
of $f_a(x)$ in $U$. This proves sufficiency.

Conversely, let $r_1,r_2,r_3\in U$ be three distinct roots
of $f_a(x)$.
Since $m$ is even, $\gcd(3,q+1)=1$, so there is a unique $\mu\in U$ satisfying $\mu^3=r_1r_2r_3$.
Then $u_i=r_i/\mu$ are distinct elements of $U$ with product $1$.
Hence
\[
\prod_{i=1}^3(x+u_i)
=x^3+\alpha x^2+\bar\alpha x+1=h_\alpha(x),
\quad\alpha=u_1+u_2+u_3.
\]
Here $\alpha\ne0$, since $x^3+1$ has only one root in $U$. It follows from Corollary~\ref{cor:deg3} that $\alpha\in\mathcal A$.
Suppose 
\[
\mu^{-5}f_a(\mu x)=h_\alpha(x)(x^2+bx+c)
\]
and comparing coefficients gives
\[
b=\alpha,\quad
c=\mu^{-5}=\frac{\alpha}{\bar\alpha},\quad\text{and}\quad
a\mu^{-2}=\bar\alpha+\alpha^2+c.
\]
The last two identities yield the stated relations. This 
proves necessity.

Finally, we determine the exact number of roots.
Since $f_a(x)$ has repeated roots if and only if $a\in U$.
For $a\in U$, however,
\[
f_a(x)=(x^2+a)(x^3+\bar a)
\]
has at most two distinct roots in $U$ since $\gcd(3,q+1)=1$.
Thus, if $f_a(x)$ has at least three distinct roots
in $U$, then it has no repeated roots.
In particular, the factors $h_\alpha(x)$ and
\[
Q_\alpha(x)=x^2+\alpha x+\frac{\alpha}{\bar\alpha}
\]
have no common root.
By Corollary~\ref{cor:deg2}, $Q_\alpha(x)$ has two distinct
roots in $U$ when $\Tr_m(1/(\alpha\bar\alpha))=1$
and none when the trace is zero.
Combining this result with the factorization of $\mu^{-5}f_a(\mu x)$
proves~(i) and~(ii).
\end{proof}

 \begin{remark}
This proposition complements the counting results in~\cite{Xia2016} .
 \end{remark}

\section{Correlation distribution for the Niho exponent $5q-1$}\label{sec:4s=5}
With the above preparations, we will investigate the correlation distribution for the Niho exponent $d$ in \eqref{ourniho} in this section.  From now on we always assume that $m\ge 2$ is even, which ensures $\gcd(d,q^2-1) = 1$.
Since $4s\equiv5\pmod{q+1}$, we have  
\(
4d\equiv5q-1\pmod{q^2-1}. 
\)
Thus, $d$ and $5q-1$ lie in the same $2$-cyclotomic coset
modulo $q^2-1$ and yield the same cross-correlation distribution
and Walsh spectrum.
We therefore use $d$ and $5q-1$ interchangeably in what follows. Using the first four power moments, we first express the
correlation distribution in terms of the three frequencies
$N_4$, $N_5$, and $N_6$
(see Proposition~\ref{thm:reconstruct}).
We then derive three identities involving these frequencies
by analyzing the associated key polynomial.

\subsection{Seven-valued correlation distribution in terms of $N_4, N_5, N_6$}

 Recall that, for the Niho exponent $d$ in \eqref{ourniho}, the cross-correlation function $C_d(\tau)$ 
 can be rewritten as $$C_d(\tau)=W_d(\alpha^{\tau})-1,\quad 0\le \tau<q^2-1,$$ where 
 $$W_d(a)=\sum_{x\in\F_{q^2}}(-1)^{\Tr_{2m}(x^d+ax)}.$$ 
 As $\tau$ ranges over $\{0,1,\cdots,q^2-2\}$, $\alpha^{\tau}$ runs through $\F_{q^2}^*$. Hence, determining the correlation distribution is equivalent to determining  the value distribution of $W_d(a)$ as $a$ ranges over $\F_{q^2}^*$.
By Niho's theorem (Lemma~\ref{lem:niho}), this reduces to counting the distinct roots in $U$ of the key polynomial
$f_{s,a}(x)=x^{2s-1}+ax^s+\bar a x^{s-1}+1$.
For $x\in U$, the congruence $4s\equiv5\pmod{q+1}$ gives $
f_{s,a}(x^4)=x^6+ax^5+\bar a x+1$. 
Since $x\mapsto x^4$ permutes $U$, $f_{s,a}(x)$ has the same
number of distinct roots in $U$ as the following hexic polynomial 
\begin{equation}\label{eq:niho-equation}
	P_a(x)=x^{6} + a x^5 + \bar{a} x + 1.
\end{equation}
Clearly for any $a\in \F_{q^2}^*$, $P_a(x)$ has at most $6$ distinct roots in $U$.
Accordingly, we define
\begin{equation}\label{eq:N_i}
N(a):=\bigl|\{u\in U:P_a(u)=0\}\bigr|~
\text{and}~
N_i:=\bigl|\{a\in\F_{q^2}^*:N(a)=i\}\bigr|, ~0\le i\le6.
\end{equation}
Niho's theorem then gives
\[
W_d(a)=q\bigl(N(a)-1\bigr).
\]
Hence, for $a\ne0$, the Walsh transform has at most seven
values $q(i-1)$, with respective frequencies $N_i$.
The corresponding cross-correlation values are $q(i-1)-1$.

The values $N_i$'s have a close connection with power moments of the Walsh transform $W_d(a)$~\cite{Niho,Helleseth1978,Katz2012}. For convenience, set $M_0(d) =q^2-1$. For $k\ge 1$, the $k$-th \textit{power moment} of the 
Walsh transform of the function $\Tr_{2m}(x^d)$ is defined as: 
\[
M_{k}(d):=\sum_{a\in \F_{q^2}} (W_d(a))^k,
\]
and it can be transformed as
\[
M_{k}(d) = \sum_{a\in \F_{q^2}} (W_d(a))^k = q^2 \sum\limits_{ \substack{x_1 + \dots + x_k = 0\\(x_1,\dots, x_k)\in \F^k_{q^2} } } (-1)^{\Tr_{2m}(x_1^d + \cdots + x_k^d)}.
\]
For any exponent $d$, one easily has 
\[
M_1(d) = q^2 \quad \text{ and }\quad  M_2(d) = q^4.
\] For $k=3$, if $\gcd(d,q^2-1)=1$, then

\[
\begin{aligned}
M_3(d)
&= q^2 \sum_{x_1,x_2\in\F_{q^2}}
(-1)^{\Tr_{2m}\left(x_1^d+x_2^d+(x_1+x_2)^d\right)} \\
&= q^2 \sum_{x\in\F_{q^2}}\sum_{y\in\F_{q^2}}
(-1)^{\Tr_{2m}\left(y^d\left((x+1)^d+x^d+1\right)\right)} \\
&= q^4 b_3.
\end{aligned}
\] where $b_3=|\{ x\in \F_{q^2}\,:\, (x+1)^d + x^d + 1 = 0\}|$. 

These identities are well-known and can be found in \cite[Proposition 3.1]{Katz2012}. The third power moment assumes that $\gcd(d,q^2-1)=1$. Under this assumption, the map $x\mapsto x^d$ permutes $\F_{q^2}$, so $W_d(0)=0$. Hence, the sums defining these moments  can be restricted to $a\in\F_{q^2}^*$.  For the Niho exponent $d$ in \eqref{ourniho}, it is coprime to $2^{2m}-1$ whenever $m$ is even. 
The following lemma evaluates $b_3$ for this Niho exponent. 

\begin{lemma}\label{lem:3rd-pm} Let $q=2^m$ with $m$ being even.  For $d=s(q-1)+1$ with $4s \equiv 5 \pmod{q+1}$, one has
    $b_3=q+12\varepsilon$, where
	$\varepsilon = 1$ if $m\equiv2\pmod4$ and $\varepsilon = 0$ if $m\equiv0\pmod4$.
\end{lemma}

\begin{proof}
	The elements $0,1$ satisfy $(x+1)^d+x^d+1=0$. 
	Raising this equation to the fourth power gives 
	\[
	   (x+1)^{5q-1} + x^{5q-1} + 1 = 0.
	\]
	For $x\ne0,1$, let
	$y=x^q$ and multiply by $x(x+1)$ the above equation. The resulting equation is
	\[
	x(y+1)^5+(x+1)y^5+x(x+1)=0,
	\] reducing to 
	\[
	y^5+xy^4+xy+x^2=(y+x)(y^4+x) = 0.
	\]
Thus, the original equation holds if and only if
$x^q=x$ or $x^{2^{m+2}}=x$.
Since $0$ and $1$ satisfy these conditions, the complete
solution set is
	\[\F_q\ \cup\ \F_{2^{\gcd(m+2,2m)}}.\]
If $m\equiv0\pmod4$, the second field is $\F_4\subseteq\F_q$. So $b_3=q$. If $m\equiv2\pmod4$, the second field is $\F_{2^4}$ and its intersection with $\F_q$ is $\F_4$, so $b_3=q+16-4=q+12$.
\end{proof}
It follows from Lemma~\ref{lem:3rd-pm} that $M_3(d) = q^4(q+12\varepsilon)$ for the Niho exponent $d$ in \eqref{ourniho}.  
The first three power moments of $W_d(a)$, together with
the sum of all frequencies,  provide the following 
four equations in $N_i$:
\begin{equation*}
\left\{ 
\begin{array}{rcl}
	M_0(d)&=&\sum\limits_{i=0}^6 N_i = q^2 - 1, \\
	 M_1(d)&=& \sum\limits_{i=0}^6 q(i-1)N_i =q^2, \\
	M_2(d) &=&  \sum\limits_{i=0}^6 (q(i-1))^2 N_i =q^4, \\
	M_3(d)&=& \sum\limits_{i=0}^6 (q(i-1))^3 N_i = q^4 (q+12\varepsilon). \\
\end{array}
\right.
\end{equation*}Solving these equations for $N_0,N_1,N_2,N_3$ gives the following formulas.
\begin{proposition}\label{thm:reconstruct} Let $q=2^m$ with $m$ being even. For the frequencies $N_i$'s defined in \eqref{eq:N_i}, we have 
	\begin{equation}\label{eq:reconstruct}
    \left\{
		\begin{array}{rcl}
			N_0&=&\frac{1}{3}(q-1-6\eps)q+N_4+4N_5+10N_6,\\
			N_1&=&\frac{1}{2}(q-1+12\eps)q-4N_4-15N_5-36N_6-1,\\
			N_2&=&(1-6\eps)q+6N_4+20N_5+45N_6,\\
			N_3&=&\frac{1}{6}(q-1+12\eps)q-4N_4-10N_5-20N_6.\\
		\end{array}
        \right.
	\end{equation}
\end{proposition}
In what follows, we aim to determine $N_4$, $N_5$, and $N_6$. To this end, we analyze the hexic polynomial $P_a(x)$ in \eqref{eq:niho-equation} to derive some identities for the frequencies $N_4, N_5, N_6$.

\subsection{Three identities for $N_4,N_5,N_6$ from characterizations of $P_a(x)$}\label{subsec:characterization}
The hexic polynomial $P_a(x)=x^{6} + a x^5 + \bar{a} x + 1$ is a normalized SCR polynomial.  In this subsection, we investigate the cases where $P_a(x)$ has $4$, $5$ or $6$ roots in $U$ as $a$ runs through $\F_{q^2}^*$, where $q=2^m$ for even $m\geq 2$ and  then obtain three identities involving $N_4$, $N_5$ and $N_6$. 

\smallskip 

We first discuss the cases where $P_a(x)$ has at least $4$ distinct roots in $U$.
\begin{proposition}\label{prop:characterization}
    Let $q=2^m$ for an even $m\geq 2$. 
    Let $S=\{u_1,u_2,u_3,u_4\}$ be a $4$-subset of $U$ and 
    define elements
    \begin{equation}\label{eq:mu_a_beta}
        \mu := (u_1u_2u_3u_4)^{\frac{1}{4}}, \quad \alpha :=\mu^{-1}\sum_{i=1}^4u_i, \quad \beta :=\mu^{-2}\sum_{1\le i<j\le4}u_i u_j.
    \end{equation}
    Then $\mu\in U$, $\alpha\in\F_{q^2}^*$, and $\beta\in\F_q$.
    For $a\in \F_{q^2}$,
     the polynomial $P_a(x)=x^6 + ax^5 + \bar{a}x + 1$
     evaluates to zero on all four elements $u_i$ in $S$ if 
     and only if the following two hold: 
     \begin{enumerate}[label=\textup{(\roman*)}]
         \item  the elements $\mu, \alpha,\beta$ defined by $u_i$'s in $S$ 
    satisfy
    \begin{equation}\label{eq:mu_alpha_beta-relation}
        (\alpha+\bar{\alpha})^2=\beta^3+\beta, \quad (\alpha+\sqrt{\beta})\mu^{-6}
=\bar{\alpha}+\sqrt{\beta};
    \end{equation}
    \item the element $a$ can be expressed by them as follows
    \begin{equation}\label{eq:mu_alpha_beta_a-relation}
        a=\mu(\alpha^2+\beta+\mu^{-6})/\alpha.
    \end{equation}
     \end{enumerate}
\end{proposition}

\begin{proof}
    We start with the necessity of the statement. 
    Suppose $P_a(x)$ evaluates to zero on the $4$-subset $S=\{u_1,u_2,u_3,u_4\}$ of $U$, i.e., $P_a(u_i)=0$ for $i=1,2,3,4$. Consider two polynomials 
    \[
    \tilde Q(x):=\prod\limits_{i=1}^4(x+u_i) \text{ and } Q(x):=\mu^{-4}\tilde Q(\mu x)=\prod\limits_{i=1}^4(x+u_i/\mu).
    \]
    By the definition of $\mu, \alpha, \beta$ in~\eqref{eq:mu_a_beta}, we have
    \[
    Q(x)=\prod\limits_{i=1}^4(x+u_i/\mu) = x^4 + \alpha x^3 + \beta x^2 + \bar{\alpha} x + 1,
    \]
    where the coefficient of $x$ in $Q(x)$ is given by 
    $$\mu^{-3}(u_1u_2u_3+u_1u_2u_4+u_1u_3u_4+u_2u_3u_4) 
    = \mu \sum_{i=1}^4 u_i^{-1} = \overline{{\mu}^{-1}} \sum_{i=1}^4 \overline{u_i} = \bar{\alpha}$$
     and the coefficient $\beta$ satisfies $$\bar{\beta} 
     = \mu^2\sum_{1\le i<j\le4}\overline{u_i u_j} = \mu^{-2}\sum_{1\le i<j\le4}{(u_i u_j)}^{-1}(u_1u_2u_3u_4)=\mu^{-2}\sum_{1\le i<j\le4}{u_i u_j}=\beta.$$
     According to the definitions of $\tilde Q(x)$ and $Q(x)$, we see that $\tilde Q(x)$ divides $P_a(x)$, which is 
    equivalent to $Q(x) \mid \mu^{-6}P_a(\mu x)$.
    Then we have the following factorization
    \[
    \mu^{-6}P_a(\mu x) = (x^4 + \alpha x^3 + \beta x^2 + \bar{\alpha} x + 1)(x^2+\gamma x+\mu^{-6}),
    \]
    where the constant $\mu^{-6}$ in the quadratic polynomial is derived from that of $\mu^{-6}P_a(\mu x)$ immediately. Meanwhile,
    comparing coefficients of $x^2,x^3,x^4$ and $x, x^5$ on two sides of the above equality, respectively, gives the following relations:
  \begin{equation}\label{eq:deg2-4}
      \beta\mu^{-6}+\bar{\alpha}\gamma +1 =0, \quad 
      \alpha \mu^{-6}+\beta \gamma+ \bar{\alpha}= 0, \quad
      \mu^{-6} +{\alpha} \gamma +\beta=0,      
  \end{equation}
    and 
    \begin{equation}\label{eq:deg1_5}
         \alpha+\gamma=a/\mu, 
         \quad 
         \bar{\alpha} \mu^{-6} + \gamma 
     = a\mu^{-5}.
    \end{equation} Note that $\alpha\neq 0$, since otherwise $Q(x)=x^4+\beta x^2+1 = (x^2+\sqrt{\beta} x + 1)^2$ would have repeated roots.
    Then equalities in \eqref{eq:deg2-4} imply $\beta^3+\beta = (\alpha +\bar{\alpha})^2$ and $(\alpha+\sqrt{\beta})\mu^{-6}=\bar{\alpha}+\sqrt{\beta}$
    in (i); and \eqref{eq:deg1_5} with the last equality in \eqref{eq:deg2-4} implies the equality $ ({\mu(\alpha^2+\beta+\mu^{-6})})/\alpha = a$ in (ii). 
   
    Conversely, suppose for a $4$-subset $S=\{u_1,u_2,u_3,u_4\}$ of $U$ the properties (i)-(ii) are satisfied. 
    Then we can define a polynomial $$Q(x) = \prod\limits_{i=1}^4(x+u_i/\mu) = x^4 + \alpha x^3 + \beta x^2 + \bar{\alpha} x + 1.$$ 
    Following from (i) and (ii) we have
     \[
     Q(x)\bigg(x^2+\frac{\mu^{-6}+\beta}{\alpha} x + \mu^{-6}\bigg) = x^6 + a\mu^{-1} x^5 + \bar{a}\mu^{-5} x + \mu^{-6}=\mu^{-6}P_a(\mu x),
     \] where $a=({\mu(\alpha^2+\beta+\mu^{-6})})/\alpha$. This implies that $Q(x)$ divides $\mu^{-6}P_a(\mu x)$, equivalently,
     $\tilde{Q}(x)=\mu^{4}Q(x/\mu)$ divides the polynomial $P_a(x) = x^6+ax^5 + \bar{a}x+1$. That is to say, 
     the elements $u_1,u_2,u_3,u_4$ in $S$ are four distinct roots of $P_a(x)$ in $U$.   
\end{proof}

\begin{remark}\label{rem:S_N456}
    According to Proposition~\ref{prop:characterization}, given a $4$-subset $S$ of $U$, when $\mu, \alpha, \beta$ defined in \eqref{eq:mu_a_beta} satisfy the equalities in \eqref{eq:mu_alpha_beta-relation}, they completely determines an element $a$ as in\eqref{eq:mu_alpha_beta_a-relation}, for which 
the polynomial $P_a(x)=x^6+ax^5 + \bar{a} x + 1$ evaluates to zero on each element of $S$. 
That is to say, the elements $a$ for which $P_a(x)$ has at least $4$ roots in $U$ are completely determined by those
 $4$-subsets $S$ of $U$ that satisfy \eqref{eq:mu_alpha_beta-relation} via the relation in \eqref{eq:mu_alpha_beta_a-relation}. 
 Define \begin{equation}\label{def4-subsetofU}
        \mathbb{S}=\bigg\{ S \subset U: |S|=4,\,\,\prod_{u \in S}(x+u) \mid P_a(x) \text{ for certain $a$ in } \F_{q^2}  \bigg\}.
 \end{equation} 
If $N(a)=i$ for $i\in\{4,5,6\}$, the roots of $P_a(x)$ determine $\binom{i}{4}$ distinct
$4$-subsets of $U$, each of which belongs to $\mathbb S$
by Proposition~\ref{prop:characterization}. Since each member of $\mathbb S$ determines a unique $a$, no $4$-subset arises from two distinct values of $a$. Hence we have 
\begin{equation}\label{eq:S-4}
   \left |\mathbb{S}\right|
 = N_4 + 5 N_5 + 15 N_6,  
\end{equation}where $N_i$ are defined in \eqref{eq:N_i}. 
\end{remark}

Next we  consider another perspective to count the $4$-subsets in $\mathbb{S}$. First, we give some auxiliary results.

\begin{lemma}\label{lem:a_Fq} Let $q=2^m$, where $m$ is even, and let $a\in\F_{q^2}$.
The polynomial $P_a(x)$ in \eqref{eq:niho-equation} has repeated roots
if and only if $a\in\F_q$, in which case it factors as
\begin{equation}\label{eq:subfactor-1}
P_a(x)=(x+1)^2(x^4+ax^3+x^2+ax+1).
\end{equation}
Furthermore,
\begin{enumerate}[label=\textup{(\roman*)}]
\item if $a=0$, then $P_0(x)$ has three distinct roots
$1,\omega,\omega^2$ in $\F_4$, each with multiplicity two,
and $1$ is its only root in $U$;

\item if $a\in\F_q^*$, then $1$ is the unique repeated root
of $P_a(x)$ and it has multiplicity exactly two.
If $\Tr_m(a^{-1})=1$, then $1$ is its only root in $U$.
If $\Tr_m(a^{-1})=0$, there exists
$t\in\F_q\setminus\{0,1\}$ such that $a=t+t^{-1}$, and then
 \[
 N(a) =
\begin{cases}
1,&\Tr_m(t)=\Tr_m(t^{-1})=0,\\
3,&\Tr_m(t)\ne\Tr_m(t^{-1}),\\
5,&\Tr_m(t)=\Tr_m(t^{-1})=1.
\end{cases}
\]
\end{enumerate}
\end{lemma}

\begin{proof}
    The proof is straightforward and placed in Appendix A for completeness.
\end{proof}
The preceding lemma gives the exact value of $N_5$.
\begin{proposition}\label{cor:N_5} Let $P_a(x)$ be the key polynomial defined in \eqref{eq:niho-equation} and $N_5$ be defined in \eqref{eq:N_i}.  If $P_a(x)$ has exactly five distinct roots in $U$,
then $a$ must lie in $\F_q$. Consequently, we have
\[
N_5=\frac{q+1+K_m(1)}{8},
\]where $K_m(\cdot)$ denotes the Kloosterman sum over $\F_q$.
\end{proposition}

\begin{proof}
    The proof is relatively simple and placed in Appendix A.
\end{proof}

\begin{lemma}\label{lem:a_U}
    For $a\in U$, the polynomial $P_a(x)$ can be factorized as
\begin{equation}\label{eq:unitfactor}
 P_a(X)=x^6+ax^5 + \bar{a} x + 1=(X+a)(X^5+a^{-1}).
\end{equation}
\begin{enumerate}[label=\textup{(\roman*)}]
\item If $m\equiv0\pmod4$, then \[
N(a)=
\begin{cases}
1,&a=1,\\
2,&a\in U\setminus\{1\}.
\end{cases}
\]
\item If $m\equiv2\pmod4$, let $\eta$ be a generator of $U$ and define  $U_5 = \{\eta^{5i}\,:\, i = 0,1,2,\dots, \frac{q-4}{5}\}$. Then, 
\[
 N(a) = \begin{cases}
     1, & \text{ if } a \in U\setminus U_5,\\
     5,& \text{ if } a=1,  \\
     6, & \text{ if } a \in U_5\setminus\{1\}.
 \end{cases}
\]
\end{enumerate}
\end{lemma}
\begin{proof}
Similarly we place the proof in Appendix A.
\end{proof}

With the above preparations, we can establish two identities involving $N_4$ and $N_6$.

\begin{theorem}\label{thm:w_i_to_N_456}
    Let $q=2^m$ for an even $m$ and $N_i$, for $i=0,1,\cdots,6$, be defined in \eqref{eq:N_i}.  Set $\varepsilon = 1$ if $m\equiv 2\pmod{4}$, and $\varepsilon = 0$ otherwise. Let $H_1=\{w\in \F_q^*: \Tr_m(1/w)=1\}$ and
     $\mathcal{W}$ be a collection of three-element subsets of $H_1$ defined by
    \begin{equation}\label{eq:W}
    \begin{array}{l}
     \mathcal{W}=\bigg\{\{w_1,w_2,w_3\}\subset H_1: w_1w_2w_3=(w_1+w_2+w_3)(w_1+w_2+w_3+1)^2\bigg\}. 
    \end{array}    
    \end{equation} Then we have 
  \begin{equation}\label{eq:W_N456}
\begin{cases}
2|\mathcal W|=N_4+4N_5+15N_6-q\varepsilon,\\
2|\mathcal W_0|=N_4,
\end{cases}
\end{equation}
     where $\mathcal{W}_{t}$ for $t=0,1$ are subsets of $\mathcal{W}$ defined by
    \begin{equation}\label{eq:W_t}
        \mathcal{W}_t = \bigg\{\{w_1,w_2,w_3\}\in \mathcal{W}: \Tr_m(\kappa) = t\bigg\}
    \end{equation}
     with
   \begin{equation}\label{eq:kappa}
    \kappa = \frac{w_1w_2+w_1w_3+w_2w_3 + (w_1+w_2+w_3)^2}{(w_1+w_2+w_3+1)^2}.\end{equation}
\end{theorem}
\begin{proof}
According to Proposition~\ref{prop:characterization} and Remark~\ref{rem:S_N456}, we see that 
$4$-subsets $S=\{u_1,u_2,u_3,u_4\}$ of $U$ satisfying \eqref{eq:mu_alpha_beta-relation} determine all elements $a\in \F_{q^2}^*$
for which $P_a(x)=x^6 + ax^5 + \bar{a}x + 1$ has at least $4$ distinct roots in $U$ through \eqref{eq:mu_alpha_beta_a-relation}.
Then we have the relation between the number of such $4$-subsets $S$ and $N_4,N_5,N_6$ as  in \eqref{eq:S-4}.

Meanwhile, the above $4$-subsets $S$ can be determined by elements $(\mu, \alpha, \beta) \in U\times \F_{q^2}^* \times \F_q$ satisfying $(\alpha+\bar{\alpha})^2=\beta^3+\beta$ and $(\alpha+\sqrt{\beta})\mu^{-6}
=\bar{\alpha}+\sqrt{\beta}$ as in \eqref{eq:mu_a_beta}, provided that the associated polynomial $Q(x)=x^4 + \alpha x^3 + \beta x^2 + \bar{\alpha} x + 1$ has four distinct roots in $U$. Below we discuss the  cases that give rise to such  $4$-subsets $S$. 

{\it Case 1:} $\beta=0$.
By $(\alpha+\bar{\alpha})^2=\beta^3+\beta$
we have $\alpha\in\F_q^*$. In this case,
\[
Q(x)=x^4+\alpha x^3+\alpha x+1
    =(x+1)^2(x^2+\alpha x+1),
\] implying that $1$ is a repeated root of $Q(x)$ and that $Q(x)$
cannot have four distinct roots in $U$. This case is thus excluded.

{\it Case 2:} $\beta=1$. 
Again we have
$\alpha\in\F_q^*$ and then $(\alpha+\sqrt{\beta})\mu^{-6}
=\bar{\alpha}+\sqrt{\beta}$ reduces to
\[
(\alpha+1)(\mu^{-6}+1)=0.
\]
If $\alpha=1$, then $\alpha+\sqrt{\beta}=0$, so the
second condition imposes no restriction on $\mu\in U$.
Otherwise, $\mu^6=1$, which forces $\mu=1$ since
$\gcd(6,q+1)=1$ for even $m$.
We consider the following two subcases. 

{\it Case 2.1:} $\alpha=\beta=1$. In this case we have
\[
\begin{aligned}
&Q(x)
=x^4+x^3+x^2+x+1
 =\frac{x^5+1}{x+1}, \\
 \text{ and }~&\mu^{-6}P_a(\mu x)
=(x^5+1)(x+\mu^{-6}), \quad a=\mu^{-5}
\end{aligned}
\]where $\mu$ is a free element in $U$. 
In this case, $a$ belongs to $U$. By Lemma~\ref{lem:a_U}, when $m\equiv 2\pmod{4}$, the normalized SCR polynomial $Q(x)=x^4+x^3+x^2+x+1$ has four distinct roots in $U$. Then, $P_a(x)$ has exactly $5$ distinct roots in $U$ if $a=1$ (equivalently, $\mu^5=1$) and 
exactly $6$ distinct roots in $U$ if $a\in U_5\setminus \{1\}$, where $U_5$ is  defined in Lemma~\ref{lem:a_U}.
Recall that $\varepsilon = 1$ if $m\equiv 2\pmod{4}$ and $\varepsilon = 0$ if $m\equiv 0\pmod{4}$. 
It follows that there are $5\varepsilon$ elements $\mu \in U$ with $\mu^5=1$ such that $N(a)=5$ and 
$(q-4)\varepsilon$ elements $\mu \in U$ with $\mu^{-5}\in U_5\setminus \{1\}$ such that $N(a)=6$. This case contributes $(q+1)\varepsilon$ $4$-subsets to $\mathbb S$ in \eqref{def4-subsetofU}.

{\it Case 2.2:} $\alpha\in\F_q^*\setminus\{1\}$.
Then $\mu=1$, $a=\alpha$, and
\[
\begin{aligned}
Q(x)=x^4+\alpha x^3+x^2+\alpha x+1,~
P_a(x)=(x+1)^2Q(x).
\end{aligned}
\] By Lemma~\ref{lem:a_Fq}, $Q(x)$ has four distinct roots in $U$
precisely when $N(\alpha)=N(a)=5$.
Proposition~\ref{cor:N_5} shows that there are exactly
$N_5=(q+1+K_m(1))/8$ such elements $\alpha\in\F_q^*$.
By Lemma~\ref{lem:a_U}, $\alpha=1$ is among them if and only if
$m\equiv2\pmod4$. Thus, after excluding $\alpha=1$, this case contributes in total $(N_5-\varepsilon)$ $4$-subsets to $\mathbb S$, and and every corresponding parameter $a$ satisfies $N(a)=5$. 

{\it Case 3.} $\beta \in \F_q\setminus\{0,1\}$. In this case, by $(\alpha+\bar{\alpha})^2=\beta^3+\beta$ we have $\alpha \not \in \F_q$, and then by $(\alpha+\sqrt{\beta})\mu^{-6}
=\bar{\alpha}+\sqrt{\beta}$ we have 
\begin{align}\label{eq:mu}
    \mu = \left(\frac{\alpha+\sqrt{\beta}}{\bar{\alpha}+\sqrt{\beta}}\right)^{\frac{1}{6}}.
\end{align}
Here $\mu$ is uniquely determined by $\alpha, \beta$. 
Furthermore, if the quartic SCR polynomial $Q(x) = x^4 + \alpha x^3 + \beta x^2 + \bar{\alpha} x + 1$
has $4$ distinct roots in $U$, denoted by $u_1,u_2,u_3,u_4$, then the set $S=\{\mu u_i: i=1,2,3,4\}$
is a $4$-subset of $U$ such that $P_a(x)$ for $a=\mu(\alpha^2+\beta+\mu^{-6})/\alpha$ evaluates to zero on $S$, 
contributing to the count of $4$-subsets on the left-hand side of~\eqref{eq:S-4}.
 
 According to Proposition~\ref{prop:resolvent}, 
the quartic SCR polynomial $Q(x)=x^4 + \alpha x^3 + \beta x^2 + \bar{\alpha} x + 1$ with $\alpha\not\in\F_q^*$ has $4$ distinct roots $u_i$'s in $U$ if and only if its cubic resolvent  $h(y)=y^3+\beta y^2+\alpha\bar{\alpha}y
      +(\alpha+\bar{\alpha})^2$ has exactly $3$ distinct roots $w_1,w_2,w_3$ in $H_1$, which can be given by
$w_1=u_1u_2+(u_1u_2)^{-1}$, $w_2=u_1u_3+(u_1u_3)^{-1}$, and $w_3=u_1u_4+(u_1u_4)^{-1}$.
Observe that when  the roots $w_1,w_2,w_2$ of $h(y)=y^3+\beta y^2+\alpha\bar{\alpha}y
      +(\alpha+\bar{\alpha})^2$ lie in $H_1$, they uniquely determine $\beta$ and the unordered pair
      $\{\alpha,\bar\alpha\}$ through Vieta's formulas:
      $$
      \beta=w_1+w_2+w_3,~~
      \alpha\bar\alpha=w_1w_2+w_1w_3+w_2w_3,~~\text{and}~~
      \alpha+\bar\alpha=\sqrt{w_1w_2w_3}.
      $$
      This means that given three distinct elements $w_i$'s in $H_1$, they can uniquely determine $\{\alpha,\bar{\alpha}\}$ and $\beta$ for which
      $h(y)=y^3+\beta y^2+\alpha\bar{\alpha}y
      +(\alpha+\bar{\alpha})^2$ satisfies $h(w_i)=0$ for $i=1,2,3$, thereby $Q(x)=x^4 + \alpha x^3 + \beta x^2 + \bar{\alpha} x + 1$ has $4$ distinct roots in $U$. If the elements $\alpha, \beta$ further satisfy
      the relation $(\alpha+\bar{\alpha})^2=\beta^3+\beta$ which corresponds to the equality $w_1w_2w_3=(w_1+w_2+w_3)(w_1+w_2+w_3+1)^2$ in the definition of $\mathcal{W}$,
      then they together with $\mu$ given by~\eqref{eq:mu} 
      contribute two $4$-subsets to $\mathbb S$ \eqref{def4-subsetofU}, namely, one by $\alpha, \beta$ and the other one by $\bar{\alpha}, \beta$. That is to say, by the definition of $\mathcal{W}$, this case contributes $2|\mathcal{W}|$ $4$-subsets to $\mathbb S$.  

Moreover, as shown in the proof of
Proposition~\ref{prop:characterization}, in this case we have the factorization
    \[\begin{aligned}
       &\mu^{-6}P_a(\mu x) 
       = Q(x)(x^2+\gamma x+\mu^{-6})
       =(x^4 + \alpha x^3 + \beta x^2 + \bar{\alpha} x + 1)(x^2+\gamma x+\mu^{-6}),
     \end{aligned}
    \]where $\gamma=\frac{\mu^{-6}+\beta}{\alpha}$ and $\mu$ is given in \eqref{eq:mu}. 
    To determine the number of roots of $P_a(x)$ in $U$, it remains
to examine the quadratic factor $x^2+\gamma x+\mu^{-6}$.
Under the assumptions of this case, it can be directly verified that $\mu^{-6}=\gamma/\bar{\gamma}.$ Then, by Corollary \ref{cor:deg2}, this quadratic equation has two distinct roots in $U$ if $\Tr_{m}(\frac{\alpha\bar \alpha+\beta^2}{1+\beta^2})=1$, and has no root in $U$ otherwise. Note that $\frac{\alpha\bar{\alpha}+\beta^2}{1+\beta^2}$
can be expressed in terms of $w_1,w_2,w_3$ and is exactly
the parameter $\kappa$ defined in \eqref{eq:kappa}. Thus, in this case, each triple $(\alpha,\beta,\mu)$
(equivalently, each pair $(\alpha,\beta)$) determined by
$\{w_1,w_2,w_3\}\in\mathcal W_0$ yields a polynomial
$P_a(x)$ with exactly four distinct roots in $U$.

For each pair $(\alpha,\beta)$ determined by
$\{w_1,w_2,w_3\}\in\mathcal W_1$, the factors $Q(x)$ and
$x^2+\gamma x+\mu^{-6}$ have four and two distinct roots
in $U$, respectively. By Lemma~\ref{lem:a_Fq}, these two factors
have at most one common root. Hence, the corresponding
polynomial $P_a(x)$ has either five or six distinct roots in $U$. More precisely, under the above conditions, $P_a(x)$ has
exactly five distinct roots in $U$ if $a\in\F_q$,
by Proposition~\ref{cor:N_5}, and exactly six if
$a\notin\F_q$, by Lemma~\ref{lem:a_Fq}.

Combining the above arguments, Case~1 contributes no
$4$-subset to $\mathbb S$ in \eqref{def4-subsetofU}, while
Cases~2 contributes $5\varepsilon + (q-4)\varepsilon + N_5 -\varepsilon = q \varepsilon + N_5$
valid $4$-subsets to $\mathbb S$, and 
Case~3 contributes $2|\mathcal{W}|$ valid $4$-subsets to $\mathbb S$.
This gives
\[
N_5+q\varepsilon+2|\mathcal{W}|=\left|\mathbb{S}\right|
=N_4 + 5N_5 + 15N_6,
\]
which  implies the desired equality in~\eqref{eq:W_N456}. 

The analysis in Cases~1--3 further shows that, the
triples $(\alpha,\beta,\mu)$ that yield polynomials $P_a(x)$
having exactly four distinct roots in $U$ are precisely the ones determined by sets $\{w_1,w_2,w_3\}\in\mathcal{W}_0$. Thus, 
\(N_4=2|\mathcal{W}_0|.\) 
\end{proof}

From the proof of Theorem~\ref{thm:w_i_to_N_456}, we have the following characterization of elements $a$'s such that $P_a(x)$ has exactly $4,5,$ or $6$ roots in $U$, respectively.
\begin{corollary}\label{cor:W_tA_t}
    Let $q=2^m$ for an even $m\geq 2$. Let $\mathcal{W}$, $\kappa$ and $\mathcal{W}_t$ be defined as in \eqref{eq:W}, \eqref{eq:kappa} and \eqref{eq:W_t}, respectively. 
   For $t=0,1$, define 
\begin{equation}\label{eq:mathcal_A} 
{\small
\mathcal{A}_t = \left\{
\left(\frac{\alpha +\sqrt{\beta}}{\bar\alpha +\sqrt{\beta}}\right)^{\frac{1}{6}} \cdot \frac{(\alpha^2+\beta+\alpha \sqrt{\beta} + 1)}{\alpha +\sqrt{\beta}}\,:\, (\alpha,\beta) \in \Omega_t\right\},}
\end{equation}
   where $\Omega_t$ consists of the pairs $(\alpha, \beta)$ obtained from  $\{w_1,w_2,w_3\}\in \mathcal{W}_t$ by setting $\beta=w_1+w_2+w_3$ and taking 
   $\alpha$ to be either root  of $X^2+\left(\sqrt{w_1w_2w_3}\right)X+\left(\sum_{1\leq i<j\leq 3}w_iw_j\right)= 0$ in $\F_{q^2}$.
Then the following statements hold: 
\begin{enumerate}[label=\textup{(\roman*)}]
	\item $N(a)=4$ if and only if $a\in\mathcal A_0$.
	Moreover, $|\mathcal A_0|=2|\mathcal W_0|$. 
	\item $N(a)=5$ if and only if
	$
	a\in\{t+t^{-1}:t,t^{-1}\in H_1\}.
$
	\item $N(a)=6$ if and only if \[
	a\in
	\begin{cases}
		\mathcal A_1\setminus\F_q,
		& m\equiv0\pmod4,\\[1mm]
		(\mathcal A_1\setminus\F_q)\cup(U_5\setminus\{1\}),
		& m\equiv2\pmod4.
	\end{cases}
	\]
\end{enumerate}
	Here,  $N(a)$ is defined in \eqref{eq:N_i}  and $U_5=\{u^5:u\in U\}$.
\end{corollary}
\begin{proof}
    The proof is placed in Appendix A.
\end{proof}
By the results of this section, determining \(N_4\) and \(N_6\) now reduces to computing the cardinalities of \(\mathcal{W}_0\) and \(\mathcal{W}_1\). The next section is devoted to this task.

\section{Determination of the correlation distribution}\label{determiningN4N6}
\renewcommand{\a}{\mathbf{a}}
\renewcommand{\b}{\mathbf{b}}
\renewcommand{\c}{\mathbf{c}}
\renewcommand{\a}{v_1}
\renewcommand{\b}{v_2}
\renewcommand{\c}{v_3}

This section will determine the cardinality of $\mathcal{W}_t$ for $t=0,1$, thereby obtaining $N_4$ and $N_6$ and completing the proof of  Theorem~\ref{thm:distribution}.  To this end, we first relate each $\mathcal W_t$ to
a set of ordered pairs $\mathcal D_t$ with simpler
defining relations, which facilitate our computations. Define 
\begin{equation}\label{eq:D}
\mathcal D=
\left\{
(x,y)\in(\F_q\setminus\{0,1\})^2:
x\ne y,~xy\ne1
\right\}.
\end{equation} 
For each $t\in \{0,1\}$, let
\begin{equation}\label{eq:Bt}
\mathcal D_t=
\left\{
(x,y)\in\mathcal D:
\Tr_m\!\left(\tfrac{1+r}{x}\right)
=\Tr_m\!\left(\tfrac{1+r}{y}\right)
=\Tr_m\!\left(\tfrac{1+r}{z}\right)=1,
\,\Tr_m(\kappa)=t
\right\}, 
\end{equation} where
\begin{equation}\label{eq:xyz-r-kappa}
z=\frac{x+y}{1+xy},~r=xyz,
~\text{and}~
\kappa=\left(xy+xz+yz+r^2\right)^2.
\end{equation}

The following lemma shows the relation between the cardinalities of $\mathcal{W}_t$ and $\mathcal{D}_t$.   

\begin{lemma}\label{wt-to-vt}
    For $t\in \{0,1\}$, let $\mathcal{W}_t$ be defined by \eqref{eq:W} and \eqref{eq:W_t}, and $\mathcal{D}_t$ be defined by \eqref{eq:D} and  \eqref{eq:Bt}. 
    Then, $|\mathcal{W}_t| = |\mathcal{D}_t|/6$ for $t=0, 1$.
\end{lemma}

\begin{proof} The proof proceeds with the defining conditions for $\mathcal{W}_t$ and $\mathcal{D}_t$, and is placed in Appendix A.
\end{proof}

By this lemma, it remains to determine
$|\mathcal D_0|$ and $|\mathcal D_1|$.

\subsection{Calculation of the cardinality of $\mathcal D_t$ by Kloosterman  sums}\label{sec:character}

In this subsection, we compute the cardinalities of \(\mathcal{D}_0\) and \(\mathcal{D}_1\).  
For each $(x,y)\in \mathcal{D}$, define 
\begin{equation}\label{eq:F}
\begin{aligned}
&F_1(x,y)=\frac{r+1}{x},~
F_2(x,y)=\frac{r+1}{y},~
F_3(x,y)=\frac{r+1}{z},
\\
&\text{and}~F_4(x,y)=\left(xy+xz+yz+r^2\right)^2=\kappa,
\end{aligned}
\end{equation}
where $z=\frac{x+y}{1+xy}$.   Let
$\chi(\cdot)=(-1)^{\Tr_m(\cdot)}$
denote the canonical additive character of $\F_q$. According to \eqref{eq:Bt}, $|\mathcal D_t|$ can be expressed  as 
\begin{equation}\label{eq:char_sum}
       |\mathcal D_t|
    = \frac{1}{16}\sum_{(x,y)\in\calD}
 \left(1-\chi(F_1)\right)\left(1-\chi(F_2)\right)\left(1-\chi(F_3)\right)
 \left(1+(-1)^t\chi(F_4)\right),
\end{equation}
where $F_i=F_i(x,y)$ and $t\in \{0,1\}$. 
Expanding the product in \eqref{eq:char_sum} yields sixteen character sums. Symmetry in
$x,y,z$ groups them into eight classes, and their relations with $|\mathcal{D}_t|$ is given below.

\begin{proposition}\label{prop:expansion}
Let $F_i(x,y)$ be the functions defined in \eqref{eq:F}, $1\le i\le4$.
For $0\le j\le3$ and $t\in\{0,1\}$, denote
\begin{equation}\label{eq:Sje}
 S_{j,t}=\sum_{(x,y)\in\calD}
 \chi\left(\sum_{i=1}^j F_i(x,y)+tF_4(x,y)\right),
\end{equation} where the notation $S_{0,t}$ is used by assuming $\sum_{i=1}^{0}F_i(x,y)=0$.
Then we have
\begin{equation}\label{eq:vtxpand}
|\mathcal D_t| =\frac1{16}\sum_{j=0}^{3}(-1)^j\binom3j
 \left(S_{j,0}+(-1)^tS_{j,1}\right),~t\in\{0,1\}.
\end{equation}
\end{proposition}
\begin{proof}
Expand the product in \eqref{eq:char_sum} using $\chi(u)\chi(v)=\chi(u+v)$.
As shown in the proof of Lemma \ref{wt-to-vt}, permutations of
$x,y,z$ induce bijections of $\mathcal V_t$, permute $F_1,F_2,F_3$, and
leave $F_4$ unchanged, where $\mathcal V_t$ is defined in Lemma \ref{wt-to-vt}. Thus, we have the following identities
\begin{enumerate}[label=\textup{(\roman*)}]
\item
$
S_{1,t}=\sum\limits_{(x,y)\in\mathcal D}\chi(F_1+t F_4)
=\sum\limits_{(x,y)\in\mathcal D}\chi(F_2+t F_4)
=\sum\limits_{(x,y)\in\mathcal D}\chi(F_3+t F_4);
$

\item $
S_{2,t}=\sum\limits_{(x,y)\in\mathcal D}\chi(F_1+F_2+t F_4)
=\sum\limits_{(x,y)\in\mathcal D}\chi(F_1+F_3+t F_4)
=\sum\limits_{(x,y)\in\mathcal D}\chi(F_2+F_3+t F_4),
$
\end{enumerate}
where $t\in \{0,1\}$.
This shows that for each $0\leq j\leq 3$, 
$j$ functions out of $F_1,F_2,F_3$ give character sums with the same value,
namely, $S_{j,0}$ without $F_4$ and $S_{j,1}$ with $F_4$, and there are $\binom{3}{j}$ such choices. Each term has
coefficient $(-1)^j$, with an additional factor $(-1)^t$ when
$F_4$ is included. Consequently,
\[
 \begin{aligned}
 &16 |\mathcal D_t|\\=&\left(S_{0,0}+(-1)^tS_{0,1}\right)
 -3\left(S_{1,0}+(-1)^tS_{1,1}\right)+3\left(S_{2,0}+(-1)^tS_{2,1}\right)
 -\left(S_{3,0}+(-1)^tS_{3,1}\right).
 \end{aligned}
\]
This yields the desired formula \eqref{eq:vtxpand}.
\end{proof}

In the sequel we shall evaluate the eight character sums
\[
S_{0,0},\ S_{1,0},\ S_{2,0},\ S_{1,1},\
S_{0,1},\ S_{2,1},\ S_{3,1},\ S_{3,0}.
\]
Four of them  will be evaluated directly, while the other four will be expressed by
three mixed Kloosterman sums. We observe that the expressions of $F_1$, $F_2$, $F_3$ and $F_4$ above 
commonly have $x+y, \,xy$. This suggests  using the following \textit{sum-product
substitution}
\begin{equation}\label{subs-sum-prod}
  (x,y)\longmapsto(s,v) \text{ with }  s=x+y,~v^2=xy,
\end{equation}
in the evaluation of $S_{j,t}$. 
Meanwhile,  the aforementioned symmetric property allows us to choose 
any $j$ of the functions $F_1,F_2,F_3$ when calculating $S_{j,t}$.
As the functions $F_3, F_1+F_2$ depend only on $x+y$ and $xy$, we will use them as representative functions in the evaluation of $S_{j,t}$.
For instance we will consider 
\[
S_{1,0}=\sum_{(x,y)\in\mathcal D}\chi(F_3), \,S_{2,0}=\sum_{(x,y)\in\mathcal D}\chi(F_1+F_2)
~\text{and}~
S_{1,1}=\sum_{(x,y)\in\mathcal D}\chi(F_3+F_4).
\]
We also need to evaluate the character sums associated with the following four functions
\[
\begin{gathered}
F_4,~ F_1+F_2+F_3,  ~F_1+F_2+F_4, ~F_1+F_2+F_3+F_4.
\end{gathered}
\] Under the sum-product substitution, each of these can be written
as a rational function of $s$ and $v$.

Let $\F_q(X,Y)$ denote the rational function field in two
indeterminates $X$ and $Y$ over $\F_q$. The following lemma provides a general method for evaluating those character sums.
\begin{lemma}\label{lem:sum-product}
Let $R(X,Y)\in\F_q(X,Y)$ and let $\mathcal D$ be defined
in \eqref{eq:D}. Suppose that, for all
$v\in\F_q\setminus\{0,1\}$ and $s\in\F_q^*$, 
$R(s,v^2)$ is well defined and satisfies
\[
 \chi\left(R(s,v^2)\right)
 =\chi\left(a s+\frac{b}{s}+c\right),
\]
where $a$, $b$, and $c$ are functions of $v$. Then,
\begin{equation}\label{eq:sumproduct}
\begin{aligned}
 &\sum_{(x,y)\in\mathcal D}\chi\bigl(R(x+y,xy)\bigr)\\
 =&
 \sum_{v\ne0,1}\chi(c)
 \left[K_m(a,b)+K_m(a,b+v)\right]
 -2\sum_{v\ne0,1}
 \chi\left(a(v^2+1)+\frac{b}{v^2+1}+c\right),
\end{aligned}
\end{equation}
where $K_m(\cdot,\cdot)$ denotes the Kloosterman sum defined in \eqref{eq: kloostermaN_sum}. 
 \end{lemma}
\begin{proof} By the relation $s=x+y$ and $v^2=xy$, each pair $(x,y)\in \mathcal D$ uniquely determines $(s,v)$.  Since $x+y\neq 0$, we have $s\neq 0$.  Conversely,
for given $s\neq 0$ and $v$, the coordinates $x,y$ are
the roots of
\[
 Z^2+sZ+v^2=0.
\] Thus the number of  pairs $(x,y)$ given by this quadratic equation is $1+\chi\left(\frac{v^2}{s^2}\right)=1+\chi\left(\frac{v}{s}\right)$. Under this correspondence, the conditions on $x,y$ defining $\mathcal D$ become
\begin{equation}\label{eq:domain}
 v\ne0,1,~s\ne0,\,v^2+1.
\end{equation}Indeed, $v\ne0$ is equivalent to $xy\neq 0$, $v\ne1$ is equivalent to
$xy\neq 1$, and $s\ne0$ is equivalent to $x \neq y$. Moreover, one of $x,y$
equals $1$ occurs when $s=v^2+1$. Therefore,
\[
\begin{aligned}
 &\sum_{(x,y)\in\mathcal D}\chi\left(R(x+y,xy)\right)
 =\sum_{v\ne0,1}\,\sum_{s\ne0,v^2+1}
 \left(1+\chi\left(\frac{v}{s}\right)\right)
 \chi\left(as+\frac{b}{s}+c\right).
\end{aligned}
\]
For fixed $v\ne0,1$, extending the inner sum to all $s\ne0$ yields 
\[
\begin{aligned}
 &\sum_{s\ne0}
 \left(1+\chi\left(\frac{v}{s}\right)\right)
 \chi\left(as+\frac{b}{s}+c\right)\\
 =&\chi(c)\left[
 \sum_{s\ne0}\chi\left(as+\frac{b}{s}\right)
 +\sum_{s\ne0}\chi\left(as+\frac{b+v}{s}\right)
 \right]\\
 =&\chi(c)
 \bigl[K_m(a,b)+K_m(a,b+v)\bigr].
\end{aligned}
\]
It remains to subtract the contribution from $s=v^2+1$. When $s=v^2+1$, $\chi\left(\frac{v}{s}\right)=\chi\left(\frac{v}{v^2+1}\right)=1$ since $\Tr_m\left(\frac{v}{v^2+1}\right)=\Tr_m\left(\frac{1}{v+1}+\frac{1}{(v+1)^2}\right)=0$. Their total contribution is
\(
 2\chi\left(a(v^2+1)+\frac{b}{v^2+1}+c\right).
\)
Subtracting this contribution and summing over $v\ne0,1$, we get the desired identity.
\end{proof}

\begin{proposition}\label{pro:fourfirstS}
For even $m\ge2$ and $q=2^m$, we have
\begin{equation}\label{eq:fourS}
\left\{\begin{aligned}
S_{0,0}&=(q-2)(q-4),\\
S_{1,0}&=2\sigma q+8,\\
S_{2,0}&=S_{1,1}=q+4-3K_m(1)+K_m(\omega)^2,
\end{aligned}\right.
\end{equation}
where $\sigma=(-1)^{m/2}$ and $\omega\in \F_q$ satisfies $\omega^2+\omega+1=0$. 
\end{proposition}
\begin{proof}
The first identity counts elements in $\calD$.  For each $x\in\F_q\setminus\{0,1\}$,
the excluded values of $y$ are $0,1,x,x^{-1}$,
which are distinct.
Hence $|\mathcal D|=(q-2)(q-4)$. The remaining evaluations can be obtained by Lemma \ref{lem:sum-product} and the results in Lemmas \ref{lem: ks_identities1} and \ref{lem: ks_identities2}. We prove the formula for $S_{1,0}$ here and give the analogous
proofs for $S_{1,1}$ and $S_{2,0}$ in Appendix B.

 For $S_{1,0}$, substituting $z=(x+y)/(xy+1)$ into $F_3(x,y)$ in \eqref{eq:F} gives
\begin{equation*}\label{F1xy}
   F_3=\frac{1+xy}{x+y}+xy. 
\end{equation*}
Using the sum-product substitution in \eqref{subs-sum-prod}, we have 
\begin{equation*}\label{F1xy}
   F_3=\frac{1+v^2}{s}+v^2~\text{and}~ \chi\left(F_3\right)=\chi\left(\frac{v^2+1}{s}+v\right).\end{equation*}
   Utilizing Lemma \ref{lem:sum-product}, we get $a=0$, $b=v^2+1$ and $c=v$ and
\[
\begin{aligned}
S_{1,0}
&=\sum_{v\ne0,1}\chi(v)\left[K_m(0,v^2+1)+K_m(0,v^2+v+1)\right]-2\sum_{v\ne0,1}\chi\left(1+v\right)\\[3pt]
&=\sum_{v\ne0,1}\chi(v)K_m(0)+\sum_{v\ne0,1}\chi(v)K_m(0)+q\chi(\omega)+q\chi(\omega^2)+4\\
&=2\times (-1)\times (-2)+2\sigma q+4\\
&=2\sigma q+8,
\end{aligned}
\]where we use the basic identity in \eqref{klkeyp1}, and $\chi(\omega)=\chi(\omega^2)=\sigma$. 
\end{proof}

To simplify the remaining four classes, we continue using the technique presented in Lemma \ref{lem:sum-product}.
The resulting expressions will be given in terms of the following three
mixed Kloosterman sums over $\F_q$:

\begin{align}
 \mathbf{K}_1&=\sum_{u\ne0}\chi(u)K_m(u+u^{-2}),\label{eq:K1}\\
 \mathbf{K}_2&=\sum_{u\ne0,1}\chi(u)
 K_m\left(\frac{(u^2+u+1)^2}{u(u+1)^2}\right),\label{eq:K2}\\
 \mathbf{K}_3&=\sum_{u\ne0,1}
 K_m\left(\frac{(u^2+u+1)^3}{u^2(u+1)^2}\right).\label{eq:K3}
\end{align}

\begin{proposition}\label{pro:remainingfour} Let $S_{j,t}$ be the character sums given in \eqref{eq:Sje}. For even $m\ge2$, we have
\begin{equation}\label{eq:Sremaining}
\left\{\begin{aligned}
 S_{0,1}&=2\sigma q+7+\mathbf{K}_1,\\
 S_{2,1}&=2\sigma q+6+\mathbf{K}_2,\\
 S_{3,1}&=2q+2-3K_m(1)+K_m(\omega)^2+\mathbf{K}_3,\\
 S_{3,0}&=2\sigma q+5+\mathbf{K}_1+\mathbf{K}_2,
\end{aligned}\right.
\end{equation}
where $\sigma=(-1)^{m/2}$ and $\omega^2+\omega+1=0$. 
\end{proposition}
\begin{proof} The equalities are derived by routine calculations. See details in Appendix B. 
\end{proof}

Proposition~\ref{pro:remainingfour} reduces the evaluation of
$S_{0,1}$, $S_{2,1}$, $S_{3,1}$, and $S_{3,0}$ to that of
the three mixed Kloosterman sums $\mathbf{K}_1$, $\mathbf{K}_2$, and $\mathbf{K}_3$. The value of $\mathbf{K}_3$ was already given in \cite[Lemmas B.3 and B.9]{XiongYan2026}, 
namely, \begin{equation}\label{valueofK3}\mathbf{K}_3=2K_m(\omega)^2-q. \end{equation}
In the following subsection we will consider $\mathbf{K}_1$ and $\mathbf{K}_2$.

\subsection{Calculations of the mixed Kloosterman sums $\mathbf{K}_1$ and $\mathbf{K}_2$}\label{sec:mixed}

This subsection will derive the following two identities: 
\begin{equation}\label{eq:intromixed}
 \mathbf{K}_1=K_m(1)^2+2\sigma q \quad\text{and}\quad
 \mathbf{K}_2=2K_m(1)^2+2(\sigma-1)q,
 \end{equation}where $\sigma=(-1)^{m/2}$.
The following result is useful for calculating $\mathbf{K}_1$. 
\begin{lemma}\cite[Propostion 3.2]{Bluher2021}\label{lem:Dickson} Let $q=2^m$, where $m\ge1$, and set $H_1=\{t\in\F_q^*:\Tr_m(t^{-1})=1\}$. For an integer $k\ge0$, let $D_k(x)$ denote the Dickson
	polynomial of the first kind with parameter $1$.
	Then $D_k(x)$ permutes $H_1$ if and only if
	$\gcd(k,q+1)=1$.
	In particular, when $m$ is even, $D_3(x)=x^3+x$
	permutes $H_1$.
\end{lemma}

\begin{proposition}\label{pro:K1} Let $\mathbf{K}_1$ be the mixed Kloosterman sum defined in \eqref{eq:K1}. 
For even $m\ge2$,
\begin{equation}\label{eq:J2value}
 \mathbf{K}_1=K_m(1)^2+2\sigma q,
\end{equation}
where $\sigma=(-1)^{m/2}$. 
\end{proposition}
\begin{proof}
Expanding the Kloosterman sum in \eqref{eq:K1} yields 
\[
 \mathbf{K}_1=\sum_{u\ne0}\sum_{c\ne0}
 \chi\left(u+c+\frac{u+u^{-2}}c\right).
\]
Let $c=v^2$. Then
\[
\chi\left(u+v^2+u/v^2+(uv)^{-2}\right)=
\chi\left ( v+(1+v^{-2})u+v^{-1}/u\right).
\]
Interchanging the sums and using identity \eqref{klkeyp1} give
\begin{equation}\label{J2newform}
 \begin{aligned}
 \mathbf{K}_1&=\sum_{v\ne0}\chi(v)
 \sum_{u\ne0}\chi\bigl((1+v^{-2})u+v^{-1}/u\bigr)\\
 &=\sum_{v\ne0}\chi(v)K_m(v^{-1}+v^{-3})\\
 &=\sum_{t\ne0}\chi(t^{-1})K_m(t^3+t)\\
 &=\sum_{t\ne0}K_m(t^3+t) - 2\sum_{t\in H_1}K_m(D_3(t)),
 \end{aligned}
\end{equation}
where $t=v^{-1}$.  Since $D_3(t)$ permutes $H_1$, Lemma \ref{lem:Dickson} gives
\[
\sum_{t\in H_1}K_m(D_3(t))
=\sum_{t\in H_1}K_m(t).
\]
By Lemma \ref{lem: ks_identities1} (\textup{i}) and Lemma \ref{lem: ks_identities2} (\textup{ii}),  the latter sum  can be evaluated as follows
\[
 \begin{aligned}
 2\sum_{t\in H_1}K_m(t)
 =&\sum_{t\ne0}\left(1-\chi(t^{-1})\right)K_m(t)\\
=&\sum_{t\ne0}K_m(t)-\sum_{t\ne0}\chi\left(t^{-1}\right)K_m(t)\\
=&1-(K_m(1)^2-q).
 \end{aligned}
\]
Then it follows from Lemma \ref{lem: ks_identities2} (\textup{iv})  and \eqref{J2newform}  that
\[
 \mathbf{K}_1=(q(2\sigma+1)+1)-(q+1-K_m(1)^2)=K_m(1)^2+2\sigma q.
\]
\end{proof}

To evaluate $\mathbf{K}_2$,  we introduce an auxiliary  character sum  $\mathbf{K}$ over a cubic surface. 
The character sum $\mathbf{K}$ can be expressed in terms of $\mathbf{K}_2$ and $\mathbf{K}_1$. Meanwhile, 
its exact value can be calculated from a different perspective. The obtained results  together with Proposition~\ref{pro:K1} enable us to determine the exactly value of $\mathbf{K}_2$.  The details are given as follows.

Let $\mathcal X $ be the cubic surface defined by 
\begin{equation}\label{eq:surface}
 \mathcal X=\{(x,y,z)\in(\F_q^*)^3:
 xyz=x+y+z\},
\end{equation}
Here, $(x,y,z)$ denotes an ordered triple, and the coordinates $x$, $y$, and $z$  are not required to be distinct. Define 
\[
 \mathbf{K}=\sum_{(x,y,z)\in\mathcal X}\chi\left((1+r)(\frac1x+\frac1y+\frac1z)\right),
\]where $r=xyz$.  Using the notation in \eqref{eq:F}, $\mathbf{K}$ can be rewritten as
\[
 \mathbf{K}=\sum_{(x,y,z)\in\mathcal X}\chi\left(F_1+F_2+F_3\right).
\]
It can be observed that  the character sum $\mathbf{K}$  is closely
related to $S_{3,0}$, but differs from it in the restrictions
on $x$, $y$, and $z$. First, we establish the following relation.

\begin{lemma}\label{prop:KK1K2}With the notation introduced above, the character sum $\mathbf{K}$ over the cubic surface $\mathcal X $ satisfies
\[
 \mathbf{K}=\mathbf{K}_2+\mathbf{K}_1+2\sigma q,
\]where $\mathbf{K}_1$ and $\mathbf{K}_2$ are defined in \eqref{eq:K1}  and \eqref{eq:K2}, respectively. 
\end{lemma}
\begin{proof}  Let $S_{3,0}$ be the character sum evaluated in Proposition~\ref{pro:remainingfour}, and $\mathcal D$ be defined in \eqref{eq:D}. 
Define 
\[
\mathcal P
=
\{(1,1,1)\}
\cup
\bigcup\limits_{t\ne0,1}
\{(1,1,t),(1,t,1),(t,1,1)\}.
\] The points of $\mathcal X$ not represented by
pairs in $\mathcal D$ are precisely those in $\mathcal P$. 
 Then,
we have 
$$\begin{aligned}\mathbf{K}
&=S_{3,0}+\sum\limits_{(x,y,z)\in \mathcal P}\chi\left(F_1+F_2+F_3\right)\\
&=S_{3,0}+\chi(0)+3\sum_{t\ne0,1}\chi\left(\frac{1}{t}\right)\\
&=S_{3,0}-5.
\end{aligned}$$
This together with the value of $S_{3,0}$ in Proposition~\ref{pro:remainingfour} gives the desired formula. 
\end{proof}

The relation between the character value and the permutation sign
used below is the cubic case of Berlekamp's discriminant
criterion \cite{Berlekamp1976}. This result is also a direct consequence of
\cite[Theorem~1]{Williams1975}. It plays an important role in the evaluation of $\mathbf{K}$.

\begin{lemma}\label{lem:cubic} Let $q=2^m$, where $m$ is a positive integer, and let
$a\in\F_q$ and $b\in\F_q^*$. Denote by $n$ the number of distinct roots of
$Z^3+aZ+b$ in $\F_q$. The number of ordered
triples $(z_1,z_2,z_3)\in\F_q^3$ satisfying
\(
 Z^3+aZ+b=(Z+z_1)(Z+z_2)(Z+z_3)
\)
is
\[
 2n+\chi\left(1+\frac{a^3}{b^2}\right)-1,
\]  where 
$\chi(\cdot)=(-1)^{\Tr_m(\cdot)}$
is the canonical additive character of $\F_q$.
\end{lemma}

Using the above result, we can now prove the following result.

\begin{proposition}\label{prop:cubiccount} Let $F(u)=u^4+u^2+u+u^{-1}$ and 
 $\mathbf{K}_4=\sum\limits_{u\ne0,1}K_m(F(u))$. Then, the character sum $\mathbf{K}$ over the cubic surface $\mathcal X$ in \eqref{eq:surface}  satisfies \[\mathbf{K}=3\mathbf{K}_4+q-3.\]
\end{proposition}
\begin{proof} We express $\mathbf{K}$ in terms of root counts of cubic
polynomials over $\F_q$.
For each point $(x,y,z)\in \mathcal X$, let
\begin{equation}\label{defrb}
 r=x+y+z,\quad b=xy+xz+yz.
\end{equation}
Since $xyz\neq 0$, it follows that $r=x+y+z=xyz\ne0$.
The coordinates $x,y,z$ are the roots, with multiplicities, of
\[
 f_{r,b}(X)=X^3+rX^2+bX+r.
\] Conversely, for any $r\ne0$ and $b\in\F_q$, every ordering
of all three roots of $f_{r,b}(X)$ in $\F_q$ gives a point of
$\mathcal X$. 

By the substitution $X=Z+r$, we get \begin{equation}\label{eq:cubic}
 f_{r,b}(Z+r)=Z^3+(b+r^2)Z+r(b+1).
\end{equation}
This translation preserves both root multiplicities and the
number of distinct roots in $\F_q$.
We can therefore count the ordered triples of roots
of $f_{r,b}(X)$ in $\F_q$ by applying
Lemma~\ref{lem:cubic} to  \eqref{eq:cubic} for $b\neq 1$. Note that the cubic polynomial  in \eqref{eq:cubic}
have no repeated roots if and only if $r(b+1)\ne0$. 

For each $r\ne0$ and $b\in\F_q$, let $n(r,b)$ be the number
of distinct roots of $f_{r,b}(X)$ in $\F_q$, and let $N(r,b)$
be the number of ordered triples $(x,y,z)\in\left(\F_q^*\right)^3$ satisfying
\[
 f_{r,b}(X)=(X+x)(X+y)(X+z).
\]
Define 
\begin{equation}\label{defWrb}
 W(r,b)=\chi\left(\frac{b(r+1)}{r}\right), ~r\neq 0, ~b\in \F_q.
\end{equation}
Then, by \eqref{defrb}, we have 
\[\mathbf{K}=\sum\limits_{r\neq 0}\,\sum\limits_{b\in \F_q}N(r,b) W(r,b).
\]
We will consider two cases: $b=1$ and $b \neq 1$.

\textbf{\textit{Case 1.}} Suppose $b=1$. Then the factorization
$f_{r,1}(X)=(X+r)(X+1)^2$ gives
\begin{equation}\label{Nrbr=1}
 (n(r,1),N(r,1))=
 \begin{cases}
 (2,3),&\text{if}~r\ne1,\\
 (1,1),&\text{if}~r=1.
 \end{cases}
\end{equation}
That is to say, $N(r,1)=2 n(r,1)-1$. Thus 
\[
\sum\limits_{r\neq 0}N(r,1) W(r,1)=N(1,1) W(1,1)+\sum\limits_{r\neq 0,1}N(r,1) W(r,1).
 \]
 According to \eqref{Nrbr=1} and \eqref{defWrb}, we obtain that 
\begin{equation}\label{Kb=1}
 \sum\limits_{r\neq 0}N(r,1) W(r,1)=3\sum_{r\ne0,1}\chi(r^{-1})+1=-5.
\end{equation}

\textbf{\textit{Case 2.}} Suppse $b\neq 1$. Then the constant term of the polynomial in \eqref{eq:cubic}  is nonzero, so
Lemma~\ref{lem:cubic}, together with $\chi(1)=1$, gives
\begin{equation}\label{Nrbrnot1}
 N(r,b)=2n(r,b)-1+
 \chi\!\left(\frac{(r^2+b)^3}{r^2(b+1)^2}\right).
\end{equation}
We need to investigate the following sum 
\begin{equation}\label{G3final}
\begin{aligned}
&\sum_{r\ne0}\sum_{b\ne1}N(r,b)W(r,b) \\
= &2\sum_{r\ne0}\sum_{b\ne1}W(r,b)n(r,b)
+\sum_{r\ne0}\sum_{b\ne1}W(r,b)
  \chi\!\left(\frac{(r^2+b)^3}{r^2(b+1)^2}\right)
-\sum_{r\ne0}\sum_{b\ne1}W(r,b).
\end{aligned}
\end{equation}
It remains to calculate each term in \eqref{G3final}. We first evaluate the three inner sums over $b\neq 1$ in \eqref{G3final} for fixed $r\ne0,1$. We then treat $r=1$, where $W(1,b)=1$,
before summing over all $r\ne0$. The computations are divided into four parts as follows. 

\medskip
\textbf{\textrm{Part (i):}}  Calculating  $\sum\limits_{b\ne1}W(r,b)\,n(r,b)$ for given $r\neq 0, 1$. 

 Now fix $r\ne0,1$. Then, we have \begin{equation}\label{fKKK0}
    \begin{aligned}
 \sum_{b\neq1}W(r,b)n(r,b)
 &=\sum_{b\in\F_q}W(r,b)n(r,b)-W(r,1)n(r,1)\\
 &=\sum_{b\in\F_q}W(r,b)n(r,b)-2\chi(1/r),\\
\end{aligned}
\end{equation}
where the second equality follows from  \eqref{Nrbr=1}. 
 
 Since $f_{r,b}(0)=r\ne0$, 
\[
 n(r,b)=|\{x\in\F_q^*:f_{r,b}(x)=0\} |.
\]
Expanding this root count and interchanging the finite sums gives
\begin{equation}\label{finialKeq1}
    \begin{aligned}
 \sum_{b\in\F_q}W(r,b)n(r,b)
 &=\sum_{b\in\F_q}
   \sum_{\substack{x\ne0\\f_{r,b}(x)=0}}W(r,b)
  =\sum_{x\ne0}
   \sum_{\substack{b\in\F_q\\f_{r,b}(x)=0}}W(r,b).
\end{aligned}
\end{equation}
For each $x\ne0$, the equation $f_{r,b}(x)=0$ determines
exactly one value of $b$, namely
\[
  b=x^2+r x+\frac r x.
\]
Thus the pairs $(b,x)$ with $f_{r,b}(x)=0$ are in bijection
with $\F_q^*$ via their second coordinate. Substitution of this value of $b$ gives
\begin{equation}\label{finialKeq2}
\begin{aligned}
 W(r,b)
 &=\chi\left(\frac{r+1}{r}x^2+(r+1)x
                   +\frac{r+1}{x}\right)
 =\chi\left(a(r)x+\frac{r+1}{x}\right),
\end{aligned}
\end{equation}
where \[
 a(r)=r+1+\sqrt{\frac{r+1}{r}}.
\] and $\sqrt{\frac{r+1}{r}}$ denotes the unique square root in $\F_q$.

Combining \eqref{finialKeq1} and \eqref{finialKeq2}, we obtain
\begin{equation}\label{finialKeq3}
\begin{aligned}
 \sum_{b\in\F_q}W(r,b)n(r,b)
 &=\sum_{x\ne0}\chi\left(a(r)x+\frac{r+1}{x}\right)\\
 &=K_m\left(a(r)(r+1)\right)\\
  &=K_m\left(a(r)^2(r+1)^2\right).
\end{aligned}
\end{equation} where the second equality holds since $r+1\neq 0$.
Note that 
\begin{align*}
 a(r)^2(r+1)^2
 &=\frac{(r+1)^3}{r}\bigl(r(r+1)+1\bigr)\\
 &=\frac{(r^3+1)(r+1)^2}{r}=F(r),
\end{align*}
where $F$ is the function appearing in the definition of $\mathbf{K}_4$.

By \eqref{fKKK0} and \eqref{finialKeq3}, we obtain 
\begin{equation}\label{eq:rootterm}
\begin{aligned}
 \sum_{b\ne1}W(r,b)n(r,b)
 &=\sum_{b\in\F_q}W(r,b)n(r,b)-W(r,1)n(r,1)\\
 &=K_m\left(F(r)\right)-2\chi(1/r).
\end{aligned}
\end{equation}

\medskip

\textbf{\textrm{Part (ii):}} Calculating $\sum\limits_{b\ne1}
 W(r,b)\chi\!\left(\frac{(r^2+b)^3}{r^2(b+1)^2}\right)$ for $r\neq 0,1$

For fixed $r\neq 0,1$, the substitution $w=b+1$ is a bijection from
$\F_q\setminus\{1\}$ to $\F_q^*$. Using this substitution, the expression $W(r,b)\chi\!\left(\frac{(r^2+b)^3}{r^2(b+1)^2}\right)$ becomes
\[
\begin{aligned}
&\chi\left(\frac{r+1}{r}(w+1)+\frac{(r^2+1+w)^3}{r^2w^2}\right)\\
 =&\chi\left(\left(\frac{r+1}{r}+\frac1{r^2}\right)w+\left(\frac{r+1}{r}+\frac {r^2+1}{r^2}\right)+\frac{(r^2+1)^2}{r^2w}+\frac{(r^2+1)^3}{r^2w^2}\right)\\
 =&\chi\left(\frac{h_r}{r^2}w+(r^{-1}+r^{-2})
 +\frac{(r+1)^4}{r^2w}
 +\left(\frac{(r+1)^3}{rw}\right)^2\right)\\
 =& \chi\!\left(\frac{h_r}{r^2}w+\frac{(r+1)^3}{r^2w}\right),
\end{aligned}
\] where $h_r=r^2+r+1$. Hence
\begin{equation}\label{eq:roottermadd2}
 \begin{aligned}
\sum\limits_{b\ne1}
 W(r,b)\chi\!\left(\frac{(r^2+b)^3}{r^2(b+1)^2}\right)
 =&\sum_{w\ne0}\chi\!\left(\frac{h_r}{r^2}w+\frac{(r+1)^3}{r^2w}\right)\\
 =&K_m\!\left(\frac{h_r(r+1)^3}{r^4}\right)\\
 =&K_m(F(r^{-1})),
\end{aligned}   
\end{equation}
where the third equality holds since $(r+1)^3/r^2\ne0$.

\medskip

\textbf{\textrm{Part (iii):}} Calculating $\sum\limits_{b\neq 1}W(r,b)$ for given $r\neq 0,1$. Note that for $r\neq 0,1$
\[
\begin{aligned}
 \sum_{b\in\F_q}W(r,b)= \sum_{b\in\F_q}\chi\left(\frac{r+1}{r}b\right)
=0,
\end{aligned}
\]
 Hence
\begin{equation}\label{eq:roottermadd1}
\begin{aligned}
 \sum_{b\ne1}W(r,b)=-W(r,1)=-\chi(1/r).
\end{aligned}
\end{equation}

\medskip

For each $r\ne0,1$, combining  \eqref{eq:rootterm},
\eqref{eq:roottermadd1}, and \eqref{eq:roottermadd2} yields
\begin{equation}\label{eq:roottermadd3}
\begin{aligned}
&\sum_{b\ne1}N(r,b)W(r,b)\\
=&2\sum\limits_{b\neq 1}W(r,b)n(r,b)-\sum\limits_{b\ne1}W(r,b)+\sum\limits_{b\ne1}
 W(r,b)\chi\!\left(\frac{(r^2+b)^3}{r^2(b+1)^2}\right)\\
 =& 2K_m\left(F(r)\right)+K_m\left(F(r^{-1})\right)-3\chi(r^{-1}).
 \end{aligned}
 \end{equation}

\medskip
\textbf{\textrm{Part (iv):}} This part is concerned with $\sum\limits_{b\ne1}N(r,b)W(r,b)$ at $r=1$. According to \eqref{G3final}, we evaluate the three sums in Parts~(i)--(iii) at $r=1$ directly from their expressions. Note that $W(1,b)=1$ for every $b\in\F_q$. The results are given as follow:

\begin{enumerate}[label=(\alph*)]
 \item 
 $ 
 \sum\limits_{b\ne1}W(1,b)n(1,b)
 =\sum\limits_{b\in\F_q}n(1,b)-n(1,1)
 =q-2
 $ since $n(1,1)=1$;

\item  $ \begin{aligned}
 \sum_{b\ne1}W(1,b)
 \chi\!\left(\frac{(1+b)^3}{(b+1)^2}\right)
 &=\sum_{b\ne1}\chi(b+1)
 &=-1;
 \end{aligned}
 $

 \item  $ \sum\limits_{b\ne1}W(1,b)=q-1.$
 
\end{enumerate}
\medskip
Combining \eqref{Kb=1}, \eqref{G3final},
\eqref{eq:roottermadd3}, and the results in Part~(iv), we obtain
\[
\begin{aligned}
\mathbf{K}
&=\sum_{r\ne0}\,\sum_{b\in\F_q}N(r,b)\,W(r,b)\\[2pt]
&=\sum_{r\ne0}\,\sum_{b\ne1}N(r,b)\,W(r,b)-5\\[2pt]
&=2\sum_{r\ne0,1}K_m\!\left(F(r)\right)
  +\sum_{r\ne0,1}K_m\!\left(F(r^{-1})\right)\\
&\quad
  -3\sum_{r\ne0,1}\chi(r^{-1})
  +\bigl[2(q-2)-(q-1)-1\bigr]-5\\[2pt]
&=3\mathbf{K}_4+q-3,
\end{aligned}
\]where we use the fact $\sum\limits_{r\ne0,1}K_m\!\left(F(r^{-1})\right)=\sum\limits_{r\ne0,1}K_m\!\left(F(r)\right)$.
\end{proof}

To evaluate $\mathbf{K}$, we need the following result.
\begin{lemma}\label{lem:K4} 
The character sum $\mathbf{K}_4$ in Proposition~\ref{prop:cubiccount}
satisfies $$\mathbf{K}_4=\mathbf{K}_1-q+1=K_m(1)^2+(2\sigma-1)q+1,$$
where $\mathbf{K}_1$ is defined in \eqref{eq:K1} and $\sigma=(-1)^{m/2}$. 
\end{lemma}
\begin{proof} See  Appendix B. 
\end{proof}

With the above preparations, we can get the exact value of $\mathbf{K}_2$
\begin{proposition}\label{pro:K2} For even $m\ge2$, the character sum  $\mathbf{K}_2$ defined in \eqref{eq:K2} can be expressed as
\begin{equation}\label{eq:JAvalue}
 \mathbf{K}_2=2K_m(1)^2+2(\sigma-1)q,
\end{equation}where $\sigma=(-1)^{m/2}$. 
\end{proposition}
\begin{proof}
The expression of $\mathbf{K}_2$ can be directly derived by combining Lemma \ref{prop:KK1K2}, Proposition \ref{prop:cubiccount} and Lemma~\ref{lem:K4}.
\end{proof}

\subsection{Proof of  Theorem~\ref{thm:distribution}}\label{sec:completion}
\begin{proof}[Proof of Theorem~\ref{thm:distribution}] Recall that $\sigma=(-1)^{m/2}$ and $\varepsilon=(1-\sigma)/2$.
By Propositions~\ref{pro:K1} and \ref{pro:K2}, we have 
\[
\mathbf{K}_1=K_m(1)^2+2\sigma q\quad\text{and}\quad
\mathbf{K}_2=2K_m(1)^2+2(\sigma-1)q.
\]
Substituting these values and the value of $\mathbf K_3$
from \eqref{valueofK3} into the formulas in
Proposition~\ref{pro:remainingfour}, we obtain
\[
\left\{
\begin{aligned}
S_{0,1}&=K_m(1)^2+4\sigma q+7,\\
S_{2,1}&=2K_m(1)^2+(4\sigma-2)q+6,\\
S_{3,1}&=q+2-3K_m(1)+3K_m(\omega)^2,\\
S_{3,0}&=3K_m(1)^2+(6\sigma-2)q+5.
\end{aligned}
\right.
\]
Combining these evaluations with
Propositions~\ref{pro:fourfirstS} and~\ref{prop:expansion}
gives the values of $|\mathcal D_0|$ and $|\mathcal D_1|$.
By Theorem~\ref{thm:w_i_to_N_456} and Lemma~\ref{wt-to-vt},
\[
N_4=\frac{|\mathcal D_0|}{3},
\quad
15N_6=\frac{|\mathcal D_1|}{3}-4N_5+q\varepsilon.
\]
Using these relations and
$N_5=(q+1+K_m(1))/8$ from Proposition~\ref{cor:N_5},
we obtain
\[
 \left\{\begin{aligned}
 48N_4&=q^2+(4\sigma-11)q+4K_m(1)^2+3K_m(1)-3K_m(\omega)^2+2,\\
 720N_6&=q^2+(9-52\sigma)q-10K_m(1)^2-45K_m(1)+9K_m(\omega)^2-44.
 \end{aligned}\right. 
\]
Substituting the values of $N_4$, $N_5$, and $N_6$
into the formulas in Proposition~\ref{thm:reconstruct}
yields $N_0$, $N_1$, $N_2$, and $N_3$.
This proves Theorem~\ref{thm:distribution}. 
\end{proof}

Note that in Theorem~\ref{thm:distribution} $N_1$ counts only nonzero parameters $a$. Since $W_d(0)=0$, the frequency of $0$ in the value distribution
of $W_d(a)$ as $a$ ranges over $\F_{q^2}$ is $N_1+1$. The following corollary shows that all seven correlation values occur for
even $m\ge6$.
\begin{corollary}
\label{cor:seven}
All seven correlation values in Theorem~\ref{thm:distribution} occur for even $m\ge6$. 
\end{corollary}
\begin{proof}
By the Kloosterman bound in Lemma~\ref{cor:Kbound}, we have
$|K_m(1)|,\,|K_m(\omega)|\le 2\sqrt q$. Then, based on the formulas in Theorem~\ref{thm:distribution}, we have 
\begin{alignat*}{2}
144N_0 &\ge 53q^2-219q-18\sqrt q-10,
&\qquad 120N_1 &\ge 44q^2-53q-30\sqrt q-101,\\[2pt]
16N_2  &\ge 3q^2-17q-2\sqrt q,
& 36N_3 &\ge 2q^2-3q-18\sqrt q-7,\\[2pt]
48N_4  &\ge q^2-27q-6\sqrt q+2,
& 720N_6 &\ge q^2-83q-90\sqrt q-44.
\end{alignat*}  

For even $m\ge8$, we have $q\ge256$ and
$\sqrt q\le q/16$, so all six lower bounds are positive. Moreover,
\[
8N_5=q+1+K_m(1)\ge(\sqrt q-1)^2>0.
\]
It remains to consider $m=6$. For $m=6$, Lemma~\ref{ks_recurrence} gives
$K_6(1)=-9$ and $K_6(\omega)=11$. Since $\sigma=-1$, we obtain
\[
(N_0,N_1,N_2,N_3,N_4,N_5,N_6)
=(1428,1606,744,234,64,7,12).
\]
Thus, all seven frequencies are strictly positive for even $m\ge6$, ensuring that all seven correlation values occur.
\end{proof}

\begin{example}
  For some small $m$ from $2$ to $10$, the frequencies $N_i$ computed with Magma
  agree with those given in Theorem~\ref{thm:distribution}. 
\begin{center}\small
\begin{tabular}{|r|rrrrrrr|}
\hline
$m$&$N_0$&$N_1$&$N_2$&$N_3$&$N_4$&$N_5$&$N_6$\\
\hline
2&0&14&0&0&0&1&0\\
4&88&89&56&20&0&2&0\\
6&1428&1606&744&234&64&7&12\\
8&23936&24227&12208&3672&1392&36&64\\
10&384660&386136&196320&58230&21560&121&1548\\
\hline
\end{tabular}
\end{center}
For $m=2$, we have $q=4$ and $d=5q-1\equiv4\pmod{15}$.
Since $4$ and $1$ lie in the same $2$-cyclotomic coset modulo $15$,
the $d$-decimated sequence is the original m-sequence. Thus, the
two correlation values are the autocorrelation values $-1$ and $15$.
For $m=4$, the exponent $5q-1=79$ is equivalent to the known
Niho exponent $3q-2=46$, since
$79\cdot46\equiv 2^6\pmod{255}.$
By Theorem~2 of \cite{Xia2016}, the corresponding cross-correlation
function is five-valued.
\end{example}

\section{Conclusion}\label{sec:conclusion}
Our main contributions in this paper are twofold. First, we study when low-degree
self-conjugate-reciprocal polynomials over $\F_{q^2}$ split in the unit circle $U$ and 
apply the result to characterize the conditions under which a hexic polynomial $P_a(x)=x^6+ax^5+\bar{a}x+1$, associated with the Niho exponent $d=s(q-1)+1$ with $4s\equiv5\pmod{q+1}$ in \eqref{ourniho}, has exactly $4, 5, 6$ roots.
Second, we determine the cross-correlation distribution of a binary m-sequence and its $d$-decimated sequence for this Niho exponent. In the calculation of  the frequencies $N_4$, $N_5$  and $N_6$, we investigate the involved character sums and transform them into expressions in terms of Kloosterman sums on two basic elements $1$ and $\omega$ in $\F_4$, which can be explicitly calculated by the well-known recurrence established by Carlitz in 1969. Niho in~\cite[p. 63]{Niho} gave a table of entries of the form $d(k)$, for Niho decimations $d$ having $k$-valued cross-correlation functions for $k\leq 8$. 
The Niho exponent $d$ considered in this paper includes,  up to equivalence, the examples listed as $61(5)$ and $23(5)$ for $n=8$, and $319(7)$ and
$347(7)$ for $n=12$, in Niho's Table~3-3.  In future, it would be interesting to consider the remaining entries in the table and to determine the value distribution for the decimation $d=5q-3$ discussed in Niho's thesis~\cite[Theorem~3-11]{Niho}.


\section{Appendices}

\subsection{Appendix A.}\label{sec:appendix-a}

\noindent\textbf{Proof of Lemma \ref{lem:a_Fq}.} When $a=0$, $P_0(x)=x^6+1=(x^3+1)^2$, and this case is trivial.
	Now we assume $a\neq 0$ and that $\xi$ is a repeated root of
	$P_a(x)$. It is clear that $P_a^\prime(x)=ax^4+\bar a$ has a unique root $x_0=(\bar a/a)^{1/4}\in U$ with multiplicity four. Substituting it  into
	$P_a(x_0)=0$ gives $x_0^6+ax_0^5+\bar{a}x_0+1=x_0^6+(ax_0^4+\bar{a})x_0+1=x_0^6+1 = 0$.
	As $m$ is even, $q+1$ is odd and is not divisible by three.
	Thus $\gcd(6,q+1)=1$, yielding $x_0=1$. The derivative equation $P_a^\prime(x_0)=P_a^\prime(1)=0$
	then gives $a=\bar a$. Hence $P_a(x)$ has  repeated
	roots if and only if $a \in\F_q$.

	When $a\in\F_q$, expanding the right-hand side of \eqref{eq:subfactor-1} verifies the identity. Let $a\in\F_q^*$, and write the quartic factor as
	$f_a(x)=x^4+ax^3+x^2+ax+1$. 
    Since $f_a'(x)=a(x+1)^2$ and $f_a(1)=1$, the polynomial $f_a(x)$
has no repeated roots. Note that $f_a(x)$ is a normalized SCR polynomial with coefficients in $\F_q$. By Proposition~\ref{prop:trace-transform},
its roots in $U$ can be determined from the roots of the
associated polynomial $g(y)$ in $\{0\}\cup H_1$.
Using \eqref{eq-trans1}, we obtain $$g(y)=(1+ay+y^2)^2.$$

The quadratic equation $y^2+ay+1=0$ has roots
in $\F_q$ if and only if $\Tr_m(a^{-1})=0$, or equivalently,
\[
a=t+t^{-1}
\quad\text{for some }t\in\F_q\setminus\{0,1\}.
\]
In this case, its two roots are $t$ and $t^{-1}$, which are distinct since $a\neq 0$.  Moreover,
\[
g(y)=(y+t)^2(y+t^{-1})^2,
\]
so each has multiplicity exactly two as a root of $g$. 
Note that when these roots belong to $H_1$, they correspond to two simple
roots of $f_a(x)$ in $U\setminus\{1\}$.
According to the definition of $H_1$,
we have
\[N(a)=
\begin{cases}
1,&\Tr_m(t)=\Tr_m(t^{-1})=0, \\
3,&\Tr_m(t)\ne\Tr_m(t^{-1}),\\
5,&\Tr_m(t)=\Tr_m(t^{-1})=1.
\end{cases}
\]
If $\Tr_m(a^{-1})=1$, then neither $y^2+ay+1$ nor
$g(y)$ has a root in $\F_q$. Hence, by
Proposition~\ref{prop:trace-transform}, $P_a(x)$ has only the root $1$ in $U$. \hfill$\square$

\medskip 

\noindent\textbf{Proof of Proposition \ref{cor:N_5}.} When $a\notin\F_q$, the polynomial $P_a(x)$ has no repeated
roots. If it has five distinct roots $u_1,\dots,u_5$ in $U$,
then, since it is monic of degree six with constant term $1$,
its sixth root is
\[
u_6=(u_1\cdots u_5)^{-1}\in U.
\]
Moreover, $u_6$ is distinct from $u_1,\dots,u_5$, since
$P_a(x)$ has no repeated roots. Thus, $P_a(x)$ cannot have
exactly five distinct roots in $U$ unless $a\in\F_q$.

By Lemma~\ref{lem:a_Fq}, $P_a(x)$ has exactly five distinct
roots in $U$ if and only if
\[
a=t+t^{-1}
\quad\text{for some }t\in\F_q\setminus\{0,1\}
\text{ with }t,t^{-1}\in H_1.
\]
In this case, $1$ is a root of multiplicity exactly two,
and the remaining four roots are distinct and belong
to $U\setminus\{1\}$.

     When $m$ is even, consider instead the set $T = \{t\in \F_q^*\setminus\{1\}: \Tr_m(1/t) = \Tr_m(t) = 1\}$.
     Then 
     \[
     |T| = \frac{1}{4}\sum_{x\in \F_q^*\setminus\{1\}}(1-(-1)^{\Tr_m(1/t)})(1-(-1)^{\Tr_m(t)})
     = \frac{q+1+K_m(1)}{4}.
     \]
Since $t,\,t^{-1}\in T$ gives a same element $a=t+t^{-1}$, we have $N_5 = \frac{q+1+K_m(1)}{8}$.    \hfill$\square$

\medskip 

\noindent\textbf{Proof of Lemma~\ref{lem:a_U}.} For $a\in U$ with $\bar{a} = a^{-1}$, 
it is straightforward that $P_a(x) = (x+a)(x^5 + a^{-1})$. Since $m$ is even, by Lemma \ref{lem:a_Fq},  the two factors have a common root if and only if $a=1$. 

If $m\equiv0\pmod4$, then $\gcd(5,q+1)=1$.
Hence, the map $u\mapsto u^5$ permutes $U$, so
$X^5=a^{-1}$ has exactly one solution in $U$.
This solution equals $a$ precisely when $a=1$,
which proves \textup{(i)}.

If $m\equiv2\pmod4$, then $5\mid(q+1)$.
The map $u\mapsto u^5$ on $U$ has kernel of size five
and image $U_5$. Consequently, $X^5=a^{-1}$ has five
distinct solutions in $U$ if $a\in U_5$, and none
otherwise. Adding the root $a$ and accounting for
the common root when $a=1$ gives \textup{(ii)}.  \hfill$\square$

\noindent\textbf{Proof of Corollary \ref{cor:W_tA_t}.} 
    As discussed in the proof of Theorem~\ref{thm:w_i_to_N_456}, for a set $\{w_1,w_2,w_3\}$ in $\mathcal{W}$, one can derive a polynomial $h(y)=(x+w_1)(x+w_2)(x+w_3)$, which by the description of $\Omega_t$ expands to $h(y)=x^3+\beta x^2 + \alpha\bar{\alpha} x + (\alpha+\bar{\alpha})^2$. 
    For an element $a$ determined by $\alpha,\beta$ as in the set $\mathcal{A}_0\cup \mathcal{A}_1$, it is routine to verify the following equality (as discussed in the proof of Proposition~\ref{prop:characterization})
    \[
     \mu^{-6}P_a(\mu x) = (x^4 + \alpha x^3 + \beta x^2 + \bar{\alpha} x + 1)(x^2+\gamma x+\mu^{-6})
    \]
    where $\mu=\left(\frac{\alpha +\sqrt{\beta}}{\bar\alpha +\sqrt{\beta}}\right)^{\frac{1}{6}}$, $\gamma = \frac{\beta+1}{\alpha+\sqrt\beta}$, and $\alpha+\sqrt\beta\ne0$ is ensured by the fact $\{w_1,w_2,w_3\}\subset H_1$ and $w_1w_2w_3=(w_1+w_2+w_3)(w_1+w_2+w_3+1)^2$.
    By Proposition~\ref{prop:resolvent} we know $Q(x)=x^4 + \alpha x^3 + \beta x^2 + \bar{\alpha} x + 1$
     has four distinct roots in $U$. We need to further discuss the quadratic polynomial $x^2+\gamma x+\mu^{-6}$.
    
   As shown in the proof of Theorem \ref{thm:w_i_to_N_456}, the quadratic factor $x^2+\gamma x+\mu^{-6}$ of $P_a(\mu x)$ has no root in $U$ when $\Tr_m(\kappa)=0$ and has two distinct roots in $U$ when  $\Tr_m(\kappa)=1$. If $a\in \mathcal{A}_0$, the quadratic factor $x^2+\gamma x+\mu^{-6}$ has no root in $U$, so thus $P_a(x)$ has exactly four distinct roots in $U$. Since the case $a\in \F_{q}^{*}$ has already been treated in 
   Lemma~\ref{lem:a_Fq}, we are only interested in the case $a\in \mathcal{A}_1\setminus \F_q$, for which $P_a(x)$ has exactly six distinct roots in $U$.
    
    Combining the above arguments with Lemmas~\ref{lem:a_Fq} and~\ref{lem:a_U}, we can derive the remaining claims.    \hfill$\square$

\medskip
\noindent\textbf{Proof of Lemma \ref{wt-to-vt}.} Let
\begin{equation}\label{eq:r-kappa}
r=v_1v_2v_3,
\quad
\kappa=\left(v_1v_2+v_1v_3+v_2v_3+r^2\right)^2, 
\end{equation} and define
\begin{equation}\label{eq:Vt}
\mathcal V_t=
\left\{
(v_1,v_2,v_3)\in (\F_q^*)^3
\ \middle|\
\begin{array}{l}
v_1+v_2+v_3=v_1v_2v_3, v_i\neq v_j, \\
\Tr_m\left(\frac{1+r}{v_i}\right)=1,
\quad 1\leq i<j\leq 3,\\
\Tr_m(\kappa)=t
\end{array}
\right\},
\end{equation} where $t=0,1$. Note that here $r\ne0$ and $r\ne 1$ since $v_i\in \F_q^{*}$ and $\Tr_m(\frac{1+r}{v_i})=1$.  We highlight two basic properties of these two sets. First, $\mathcal V_t$ is invariant under permutations
of the coordinates. Second, for any \((v_1,v_2,v_3)\in\mathcal{V}_t\), the relation \(v_1+v_2+v_3=v_1v_2v_3\) and the pairwise distinctness of the coordinates imply that
\[
v_i\ne1\quad(1\le i\le3),
~\text{and}~
v_iv_j\ne1\quad(1\le i<j\le3).
\]

We first establish a relation between $\mathcal{W}_t$ and the triple set $\mathcal{V}_t$ for each $t=0,1$, and then prove that $|\mathcal{W}_t|=|\mathcal{V}_t|/6$. The defining relations of $\mathcal{V}_t$, $t=0,1$,  are simpler and facilitate subsequent computations.

Suppose $\{w_1,w_2,w_3\}\in \mathcal{W}_t$. Then, they  are three distinct elements in $H_1=\{x\in \F_{q}^*: \Tr_m(1/x)=1\}$ and satisfy 
\begin{equation}\label{conditiononw3}
  w_1w_2w_3=(w_1+w_2+w_3)(w_1+w_2+w_3+1)^2.  
\end{equation}
We denote \begin{equation}\label{w-r-v}v_i=\sqrt{\frac{w_i}{w_1+w_2+w_3+1}}=\sqrt{\frac{w_i}{\beta+1}}\quad\text{for}\quad i=1,2,3,\end{equation} where $\beta=w_1+w_2+w_3$. In this way, the condition \eqref{conditiononw3} in $\mathcal{W}_t$ reduced to 
\[
(v_1v_2v_3)^2
= 
(v_1+v_2+v_3)^2(v_1^2+v_2^2+v_3^2+\frac{1}{\beta+1})^2=(v_1+v_2+v_3)^2
\] since $v_1^2+v_2^2+v_3^2= \frac{\beta}{\beta+1}$. Note that the map $\beta\mapsto\frac{\beta}{\beta+1}$ is an involution on $\F_{q^2}\setminus \{0,1\}$. The identity $v_1^2+v_2^2+v_3^2= \frac{\beta}{\beta+1}$ also implies that \begin{equation}\label{betathroughv}
  \beta+1=\frac{1}{v_1^2+v_2^2+v_3^2+1}=\frac{1}{(v_1v_2v_3)^2+1}.  
\end{equation}
In addition, the element $\kappa$ in \eqref{eq:kappa} can be expressed by $v_1,v_2,v_3$ as
$$
\kappa=\left(v_1v_2+v_1v_3+v_2v_3+(v_1v_2v_3)^2\right)^2.
$$

The relation in \eqref{w-r-v} between $v_i$ and $w_i$ gives   a one-to-one correspondence between
the ordered triples $(w_1,w_2,w_3)$ and $(v_1,v_2,v_3)$.  The condition $w_1w_2w_3=(w_1+w_2+w_3)(w_1+w_2+w_3+1)^2$ corresponds to 
a simpler condition $v_1v_2v_3 = v_1 + v_2 + v_3$. Meanwhile, the elements in a set $\{w_1,w_2,w_3\}$ are different, indicating that $v_i\neq v_j$ for $1\leq i<j\leq 3$; and $w_i\in H_1$ implies that $1=\Tr_m(\frac{1}{w_i})=\Tr_m(\frac{1}{(\beta^2+1)v_i^2})=\Tr_m(\frac{1+v_1v_2v_3}{v_i})$ due to \eqref{betathroughv}. Hence the conditions on $v_i$'s exactly correspond to those on $w_i$'s. 
Lastly, each set $\{w_1,w_2,w_3\}$ in $\mathcal{W}_t$ provides $6$ different ordered triples, implying that $|\mathcal{W}_t| = |\mathcal{V}_t|/6$. 

Next we show $|\mathcal{V}_t|=|\mathcal{D}_t|$ for $t\in\{0,1\}$. Note that each pair $(x,y)$ in $\mathcal{D}$ uniquely determines $z$ via $z=\frac{x+y}{1+xy}$,  and $(x+1)(y+1)(x+y)(xy+1)\neq 0$ ensures that $x,y,z$ are nonzero and pairwise distinct.  Thus, for every pair in $\mathcal D$, the resulting coordinates
$x,\,y,\,z=(x+y)/(1+xy)$ are nonzero and pairwise distinct, and satisfy
$x+y+z=xyz$. Moreover, $(x,y)\in\mathcal D_t$ if and only if
$(x,y,z)\in\mathcal V_t$.
Therefore, there is a one-to-one correspondence between
$\mathcal D_t$ and $\mathcal V_t$, i.e., $
|\mathcal V_t|=|\mathcal D_t|$.
 \hfill$\square$

\subsection{Appendix B.} \label{sec:appendix-b}

\noindent\textbf{Proof of Lemma \ref{lem: ks_identities2}.} 
We first derive a product identity. For a given $y\in\mathbb{F}_{q}^*$, making the substitution $x=ry$ in the expansion of $K_m(a)K_m(b)$ yields
\begin{equation}\label{trans-kaKb}
\begin{aligned}
K_m(a)K_m(b)
&=\sum_{x\in\F_q^*}\sum_{y\in\F_q^*}
  \chi\left(x+\frac{a}{x}+y+\frac{b}{y}\right)\\
&=\sum_{r\in\F_q^*}\sum_{y\in\F_q^*}
  \chi\left((r+1)y+\frac{a/r+b}{y}\right)\\
&=\sum_{y\in\F_q^*}
  \chi\left(\frac{a+b}{y}\right)+
\sum_{r\in\F_q\setminus\{0,1\}}
  K_m\left((r+1)(a/r+b)\right)\\
  &=\sum_{y\in\F_q^*}
  \chi\left(\frac{a+b}{y}\right)+
\sum_{r\in\F_q\setminus\{0,1\}}
  K_m\left(a+b + a/r+br)\right),
\end{aligned}
\end{equation}
where the term with $r=1$ is $q-1$ when $a=b$
and $-1$ otherwise, and for the terms for $r\neq 0, 1$ we use the identity \eqref{klkeyp1}.

 Taking $a=\omega$ and $b=\omega^2$ in \eqref{trans-kaKb}, and then setting
$v=\omega^2r$, gives
\[
\begin{aligned}
K_m(\omega)^2
&=-1+\sum_{r\in\F_q\setminus\{0,1\}}
  K_m(1+\omega/r+\omega^2r)\\
&=-1+\sum_{\substack{v\in\F_q^*\\v\ne\omega^2}}
  K_m(1+v+v^{-1})\\
&=\sum_{v\in\F_q^*}K_m(1+v+v^{-1}),
\end{aligned}
\]
where in the second equality we use the substitution $v=\omega^2 r$ and the last equality follows from
$1+\omega^2+\omega=0$ and $K_m(0)=-1$.
This proves \textup{(i)}.

 Taking $a=b=1$ in \eqref{trans-kaKb} and using $K_m(0)=-1$, we obtain
\begin{equation}\label{eq:K-square-pairing}
\begin{aligned}
K_m(1)^2
&=q-1+\sum_{r\in\F_q\setminus\{0,1\}}
  K_m(r+r^{-1})
  =q+\sum_{r\in\F_q^*}K_m(r+r^{-1}).
\end{aligned}
\end{equation}
We next count the preimages of the map
$v\mapsto v+v^{-1}$ on $\F_q^*$.
The value $0$ has the unique preimage $v=1$.
For $t\ne0$, the equation $v+v^{-1}=t$ is equivalent to
\[
v^2+tv+1=0.
\]
The number
of its solutions in $\F_q^*$ is $1+\chi(t^{-1})$.
Thus, for every $b\in\F_q$,
\[
\begin{aligned}
\sum_{v\in\F_q^*}K_m(b+v+v^{-1})
&=K_m(b)+\sum_{t\in\F_q^*}
  \bigl(1+\chi(t^{-1})\bigr)K_m(b+t)\\
&=\sum_{t\in\F_q^*}\chi(t^{-1})K_m(b+t),
\end{aligned}
\]
since
\[
K_m(b)+\sum_{t\in\F_q^*}K_m(b+t)
=\sum_{c\in\F_q}K_m(c)=0.
\]
Taking $b=0$ and using \eqref{eq:K-square-pairing}
proves \textup{(ii)}.
Taking $b=1$ and using \textup{(i)} gives \textup{(iii)}

Finally,  expanding the Kloosterman sum $K_m(c^3+c)$ and substituting $x=cv$ in the inner sum, we obtain
\[
 \begin{aligned}
 &\sum\limits_{c\in\F_q^*}K_m(c^3+c)\\
 =&\sum_{c\ne0}\sum_{x\ne0}\chi\left(x+\frac{c^3+c}{x}\right)\\
  =&\sum_{v\ne0}\sum_{c\ne0}\chi\left(cv+\frac{c^2}{v}+v^{-1}\right)\\
   =&\sum_{v\ne0}\chi\left(v^{-1}\right)
   \sum_{c\ne0}\chi\left(c(v+v^{-1/2})\right)\\
  =&q\sum_{v^3=1}\chi\left(v^{-1}\right)-\sum_{v\ne0}\chi\left(v^{-1}\right),
 \end{aligned}
\]
where we use the fact $v+v^{-1/2}=0$ if and only if $ v^3=1$. Since $m$ is even, the three roots are $v=1,\omega,\omega^2$, for which
$\chi(v^{-1})$ takes the values $1,\sigma,\sigma$,
respectively. Therefore, $$\sum\limits_{c\in\F_q^*}K_m(c^3+c)=(2\sigma+1)q+1.$$ 
\hfill$\square$

\medskip

\noindent\textbf{Proof of Proposition \ref{pro:fourfirstS}.} We prove the formulas for $S_{1,1}$ and $S_{2,0}$ in two parts, treating $S_{1,1}$ first and $S_{2,0}$ second.

\smallskip
  \textbf{\textrm{Part (i):}} For $S_{1,1}$, from \eqref{eq:F}, we have 
\[ 
  \chi\left (F_3+F_4\right)
 = \chi\left(\frac{1+xy}{x+y}+\frac{(x+y)(x+y)}{xy+1}+xy\frac{(x+y)}{xy+1}\right)\\
 \]
Under the sum-product substitution in \eqref{subs-sum-prod}, 
\[\begin{aligned}
  \chi\left (F_3+F_4\right)
  &=\chi\left(\frac{1+v^2}{s}+\frac{s^2}{v^2+1}+v^2\frac{s}{1+v^2}\right)\\
  &=\chi\left(\frac{1+v^2}{s}+\frac{s}{v+1}+v^2\frac{s}{1+v^2}\right)\\
   &=\chi\left(\frac{v^2+v+1}{v^2+1}s+\frac{1+v^2}{s}\right).
  \end{aligned} 
 \]
By Lemma  \ref{lem:sum-product}, we get $a=\frac{v^2+v+1}{v^2+1}$, $b=v^2+1$ and $c=0$ and
\[
\begin{aligned}
S_{1,1}
&=\sum_{v\ne0,1}\left[K_m\left(\frac{v^2+v+1}{v^2+1},v^2+1\right)+K_m\left(\frac{v^2+v+1}{v^2+1},v^2+v+1\right)\right]-2\sum_{v\ne0,1}\chi\left(v^2+v\right)\\[3pt]
&=\sum_{v\ne0,1}K_m\left(v^2+v+1\right)+\sum_{v\ne0,1}K_m\left(\frac{(v^2+v+1)^2}{v^2+1}\right)+2q-2(q-2)\\
&=\sum_{v\ne0,1}K_m\left(v^2+v+1\right)+\sum_{v\ne0,1}K_m\left(\frac{v^2+v+1}{v+1}\right)+4,
\end{aligned}
\]
where the second equality holds since $K_m\left(\frac{v^2+v+1}{v^2+1},v^2+v+1\right)=q-1$ if and only if $v^2+v+1=0$. Furthermore, by making a substitution $t=v^2+v$ in the first sum and replacing  $v$ with $v+1$ in the second sum, we get 
\[
\begin{aligned}
S_{1,1}
&=\sum_{t\ne0}\left(1+\chi(t)\right)K_m\left(t+1\right)+\sum_{v\ne0,1}K_m\left(\frac{v^2+v+1}{v}\right)+4\\
&=\sum_{t\ne0}K_m\left(t+1\right)+\sum_{t\ne0}\chi(t)K_m\left(t+1\right)+\sum_{v\ne0}K_m\left(v+v^{-1}+1\right)-K_m(1)+4\\
&=\sum_{t\in \F_q}K_m\left(t+1\right)+\sum_{t\in \F_q}\chi(t)K_m\left(t+1\right)-2K_m(1)+K_m(\omega)^2-K_m(1)+4\\
&=\sum_{t\in \F_q}\chi(t)K_m\left(t\right)+K_m(\omega)^2-3K_m(1)+4\\
&=q+K_m(\omega)^2-3K_m(1)+4,
\end{aligned}
\]
where we use the identities in Lemma \ref{lem: ks_identities1} (iii)  and Lemma \ref{lem: ks_identities2} (i) to simplify the calculation.

\smallskip
 \textbf{\textrm{Part (ii):}} For $S_{2,0}$, since $$F_1+F_2=\frac1x+\frac{y(x+y)}{xy+1}+\frac1y+\frac{x(x+y)}{xy+1}=\frac{x+y}{xy}+\frac{(x+y)^2}{xy+1}.$$
Using the substitution in \eqref{subs-sum-prod}, we get 
\[
 F_1+F_2= \frac{s^2}{v^2+1}+\frac{s}{v^2}= \frac{s^2}{(v+1)^2}+\frac{s}{v^2}.
\]
and 
\[
 \chi\left(F_1+F_2\right)=\chi\left(a s\right),\quad a=\frac{v^2+v+1}{v^2(v+1)}.
\]
By Lemma  \ref{lem:sum-product}, we have $a=\frac{v^2+v+1}{v^2(v+1)}$, $b=0$ and $c=0$.
Then, 
\begin{equation}\label{S20eq1}
    \begin{aligned}
 S_{2,0}
 &=\sum_{v\ne0,1}K_m\left(\frac{v^2+v+1}{v^2(v+1)},0\right)+\sum_{v\ne0,1}K_m\left(\frac{v^2+v+1}{v^2(v+1)},v\right)-2\sum_{v\ne0,1}\chi\left(\frac{v^3+1}{v^2}\right)\\
 &=2q+\sum_{v\ne0,1}K_m(0)+\sum_{v\ne0,1}K_m\left(\frac{v^2+v+1}{v(v+1)}\right)-2\sum_{v\ne0,1}\chi\left(v+\frac{1}{v}\right)\\
 &=2q-(q-2)+
   \sum_{v\ne0,1}K_m\left(1+\frac1{v^2+v}\right)-2K_m(1)+2\\
   &=q+4-2K_m(1)+
   \sum_{v\ne0,1}K_m\left(1+\frac1{v^2+v}\right).
 \end{aligned}
\end{equation}

Finally, let $t=v^2+v$. Then, 
\begin{equation}\label{S20eq2}
 \begin{aligned}
 \sum_{v\ne0,1}K_m\left(1+\frac1{v^2+v}\right)
 &=\sum_{t\ne0}(1+\chi(t))K_m(1+t^{-1})\\
 &=\sum_{t\ne0}K_m(1+t)
   +\sum_{t\ne0}\chi(t^{-1})K_m(1+t)\\
 &=-K_m(1)+K_m(\omega)^2.
 \end{aligned}
\end{equation}
The second equality follows by replacing \(t\) with \(t^{-1}\), and the last follows from Lemma~\ref{lem: ks_identities1} (i) and Lemma \ref{lem: ks_identities2} (iii). Combining \eqref{S20eq1} and \eqref{S20eq2}, we obtain
\[
S_{2,0}=q+4-3K_m(1)+K_m(\omega)^2.
\] \hfill$\square$
\smallskip

\noindent\textbf{Proof of Proposition \ref{pro:remainingfour}.} For the remaining four classes, indexed by
\((j,t)\in\{(0,1),(2,1),(3,0),(3,1)\}\),
we apply Lemma~\ref{lem:sum-product} by determining the coefficients
\(a\), \(b\), and \(c\), together with the value of
\(\chi\!\left(as+\frac{b}{s}+c\right)\) at \(s=v^2+1\).
\begin{table}[htbp]
\centering
\caption{Coefficients in $S_{j,t}$'s and the values of \(\chi\!\left(as+\frac{b}{s}+c\right)\) at $s=v^2+1$}\vspace{2mm}
\label{tab:remaining-classes}
\renewcommand{\arraystretch}{1.2}
\begin{tabular}{|c|ccc|c|}
\hline
Class & $a$ & $b$ & $c$
& \shortstack{$\chi\!\left(as+\frac{b}{s}+c\right)$
              at $s=v^2+1$}\\
\hline 
$S_{0,1}$
& $h_v/(v+1)^2$
& $0$ & $v$ & $\chi(v)$\\
$S_{2,1}$
& $h_v^2/(v^2(v+1)^2)$
& $0$ & $v$ & $\chi(v^{-1})$\\
$S_{3,0}$
& $h_v/(v^2(v+1))$
& $v^2+1$ & $v$ & $\chi(v^{-1})$\\
$S_{3,1}$
& $h_v^2/(v^2(v+1)^2)$
& $v^2+1$ & $0$ & $\chi(v+v^{-1})$\\
\hline 
\end{tabular}
\end{table}
The results, obtained by direct calculation, are listed in
Table~\ref{tab:remaining-classes}, where \(h_v=v^2+v+1\). Note that in each row, $a=0$ if and only if
$v\in\{\omega,\omega^2\}$, where $\omega$ is a primitive element of $\F_4$. 

Using the values in Table~\ref{tab:remaining-classes},
we apply Lemma  ~\ref{lem:sum-product} to each of the four classes in turn.  The proof are divided into four parts:
Part~(i) for  $S_{0,1}$, Part~(ii) for $S_{2,1}$, Part~(iii) for $S_{3,0}$,
and Part~(iv) for $S_{3,1}$.

\textbf{\textrm{Part (i):}} For $S_{0,1}$,  by  Table~\ref{tab:remaining-classes}, we have 
\[
 \begin{aligned}
 S_{0,1}
 &=\sum_{v\ne0,1}\chi(v)
   \left[K_m\left(\frac{h_v}{(v+1)^2}, 0\right)+K_m\left(\frac{h_v}{(v+1)^2},v\right)\right]
   -2\sum_{v\ne0,1}\chi(v)\\
&=2\sigma q+\sum_{v\ne0,1}\chi(v)
   K_m\left(0\right)+\sum_{v\ne0,1}\chi(v) K_m\left(\frac{vh_v}{(v+1)^2}\right)+4\\
 &=2\sigma q+6+
   \sum_{v\ne0,1}\chi(v)K_m\bigl(v+1+(v+1)^{-2}\bigr)\\
   &=2\sigma q+6+
   \sum_{u\ne0,1}\chi(u)K_m\bigl(u+u^{-2}\bigr),\\
    &=2\sigma q+7+\sum_{u\ne0}\chi(u)K_m\bigl(u+u^{-2}\bigr),
 \end{aligned}
\]
where we use the substitution $u=v+1$, which permutes $\F_q\setminus\{0,1\}$ and
$\chi(v)=\chi(u)$ since $\chi(1)=1$. Hence $S_{0,1}=2\sigma q+7+\mathbf{K}_1$.

\textbf{\textrm{Part (ii):}}  For $S_{2,1}$, we have 
\[
 \begin{aligned}
 S_{2,1}
 &= \sum_{v\ne0,1}\chi(v)\left[K_m\left(\frac{h_v^2}{v^2(v+1)^2}, 0\right)+K_m\left(\frac{h_v^2}{v^2(v+1)^2},v\right)\right]
   -2\sum_{v\ne0,1}\chi(v^{-1})\\
   &= 2\sigma q+\sum_{v\ne0,1}\chi(v)K_m\left(0\right)+\sum_{v\ne0,1}\chi(v)K_m\left(\frac{h_v^2}{v(v+1)^2}\right)+4\\
&=2\sigma q+6+\mathbf{K}_2.
 \end{aligned}
\]

\textbf{\textrm{Part (iii):}} For $S_{3,0}$, we have 
\[
\begin{aligned}
S_{3,0}
&=\sum_{v\ne0,1}\chi(v)\left[K_m\left(\frac{h_v}{v^2(v+1)}, v^2+1\right)+K_m\left(\frac{h_v}{v^2(v+1)},v^2+v+1\right)\right]
   -2\sum_{v\ne0,1}\chi(v^{-1})\\[3pt]
&=2\sigma q+4
 +\sum_{v\ne0,1}\chi(v)
 \left[
 K_m(v+v^{-2})
 +K_m\!\left(\frac{h_v^2}{v^2(v+1)}\right)
 \right]\\
&=2\sigma q+4+\mathbf{K}_1+1
 +\sum_{v\ne0,1}\chi(v)
 K_m\!\left(\frac{h_v^2}{v^2(v+1)}\right)\\
 &=2\sigma q+5+\mathbf{K}_1
 +\sum_{u\ne0,1}\chi(u)
 K_m\!\left(\frac{h_u^2}{u(u+1)^2}\right)
 \\[3pt]
&=2\sigma q+5+\mathbf{K}_1+\mathbf{K}_2,
\end{aligned}
\]
where the last second equality holds by the substitution $u=v+1$ and  the fact $\chi(u+1)=\chi(u)$.

\textbf{\textrm{Part (iv):}} For $S_{3,1}$, by Table~\ref{tab:remaining-classes} and  Lemma  ~\ref{lem:sum-product}, we have
\[
\begin{aligned}
S_{3,1}
&=\sum_{v\ne0,1}\left[K_m\left(\frac{h_v^2}{v^2(v+1)^2}, v^2+1\right)+K_m\left(\frac{h_v^2}{v^2(v+1)^2},v^2+v+1\right)\right]
   -2\sum_{v\ne0,1}\chi(v+v^{-1})\\
&=\sum_{v\ne0,1}K_m\left(\frac{h_v^2}{v^2}\right)+2q+\sum_{v\ne0,1}K_m\left(\frac{h_v^3}{v^2(v+1)^2}\right)-2K_m(1)+2\\
&=\sum_{v\ne0,1}K_m\left(v+v^{-1}+1\right)+2q+\mathbf{K}_3-2K_m(1)+2\\
&=\sum_{v\ne0}K_m\left(v+v^{-1}+1\right)+2q+\mathbf{K}_3-3K_m(1)+2\\
&=2q+\mathbf{K}_3+K_m(\omega)^2-3K_m(1)+2,
\end{aligned}
\]where the last equality follows from Lemma \ref{lem: ks_identities2} (i).
 \hfill$\square$

\smallskip
\noindent\textbf{Proof of Lemma \ref{lem:K4}.} 
Note that $F(u)$ has the following factorization
\[
 F(u)=(u+1)^2(u^2+u^{-1}),
\]
which suggests $K_m(F(u))$ can be written in the form $=K_m\left((u+1)^2,u^2+u^{-1}\right)$.  
For $u\ne0,1$, we make a new substitution $x=(u+1)^2t^2$ in $K_m(F(u))$.
This is a bijection of $\F_q^*$, and
\[
 \begin{aligned}
K_m(F(u))&= \sum\limits_{x\neq 0}\chi \left(x+\frac{F(u)}x\right)\\
 &=\sum\limits_{t\neq 0}\chi \left((u+1)^2t^2+\frac{u^2+u^{-1}}{t^2}\right)\\
 &=\sum\limits_{t\neq 0}\chi \left( (u+1)t+\frac ut+\frac1{ut^2}\right)\\
 &=\sum\limits_{t\neq 0}\chi \left( t+u(t+t^{-1})+\frac1{ut^2}\right).
 \end{aligned}
\]
Thus, 
\[\begin{aligned}
\mathbf{K}_4&=\sum\limits_{u\ne0}\sum\limits_{t\ne0}
 \chi\left(t+u(t+t^{-1})+\frac1{ut^2}\right)-\sum\limits_{t\ne0}\chi(0)\\
 &=\sum\limits_{t\ne0}\chi(t)\sum\limits_{u\ne0}
 \chi\left(u(t+t^{-1})+\frac1{ut^2}\right)-(q-1)\\
  &=\sum_{t\ne0}\chi(t)K_m(t^{-1}+t^{-3})-(q-1)\\
   &=\sum_{t\ne0}\chi(t^{-1})K_m(t+t^3)-(q-1).
  \end{aligned}
\]
The last expression $\sum_{t\ne0}\chi(t^{-1})K_m(t+t^3)$
for $\mathbf{K}_1$ was established in \eqref{J2newform} in the proof of Proposition  \ref{pro:K1}. 
Therefore, $\mathbf{K}_4=\mathbf{K}_1-q+1$, and \eqref{eq:J2value} gives the stated value.
 \hfill$\square$

\end{document}